\documentclass[10pt,reprint,twocolumn,prd,aps,amsmath,amssymb,nofootinbib]{revtex4-1}

\usepackage{amsfonts}
\usepackage{textcomp}
\usepackage{gensymb}
\usepackage{graphicx}
\usepackage{needspace}
\usepackage{hyperref}

\begin{document}

\title{Effects of Inner Alfv\'{e}n Surface Location on Black Hole Energy Extraction in the Limit of a Force-Free Magnetosphere}

\author{Kevin Thoelecke}
\affiliation{Department of Physics, Montana State University, Bozeman, Montana 59717, USA}

\author{Masaaki Takahashi}
\affiliation{Department of Physics and Astronomy, Aichi University of Education, Kariya, Aichi 448-8542, Japan}

\author{Sachiko Tsuruta}
\affiliation{Department of Physics, Montana State University, Bozeman, Montana 59717, USA}

\begin{abstract} 
An energy extracting black hole magnetosphere can be defined by the location of its inner Alfv\'{e}n surface, which determines the rate of black hole energy extraction along a given magnetic field line.  We study how the location of the inner Alfv\'{e}n surface can modify the structure of energy extracting black hole magnetospheres in the force-free limit.  Hundreds of magnetospheres are numerically computed via a general relativistic extension of the Newtonian magnetofrictional method for a full range of black hole spins and flow parameters.  We find that jet-like structures naturally form very close to the horizon for Alfv\'{e}n surfaces near the boundary of the ergosphere and that energy is extracted towards the equatorial plane for Alfv\'{e}n surfaces close to the horizon.  This suggests that two broad classes of energy extracting black hole magnetospheres might exist; those that transmit extracted energy directly to distant observers, and those that transmit extracted energy to nearby accreting matter.  Applied to transient high energy phenomena, we find that they might also differ in timescale by a factor of 20 or more.
\end{abstract}

\maketitle

%%%%%%%%%%%%%%%%%%%%%%%%%%%%%%%%%%%%%%%%%%%%%%%%%%%%%%%%%%%%%%%%%%%%%%%%%%%%%%%%%%%%%%%%%%%%%%%%%%%%%%%%%%%%%%%%%%%%%%%%%%%%%%%%%%%%%%
%%%%%%%%%%%%%%%%%%%%%%%%%%%%%%%%%%%%%%%%%%%%%%%%%%%%%%%%%%%%%%%%%%%%%%%%%%%%%%%%%%%%%%%%%%%%%%%%%%%%%%%%%%%%%%%%%%%%%%%%%%%%%%%%%%%%%%
%%%%%%%%%%%%%%%%%%%%%%%%%%%%%%%%%%%%%%%%%%%%%%%%%%%%%%%%%%%%%%%%%%%%%%%%%%%%%%%%%%%%%%%%%%%%%%%%%%%%%%%%%%%%%%%%%%%%%%%%%%%%%%%%%%%%%%
\section{Introduction}

When a rotating black hole is immersed in a magnetic field supported by a plasma, it is possible for that magnetic field to carry an electromagnetic Poynting flux of energy and angular momentum away from the black hole to a distant observer.  Conservation of energy and angular momentum requires that the spin of the black hole be reduced; it is therefore said that such a magnetosphere extracts rotational energy from the black hole.

This type of electromagnetic black hole energy extraction is often referred to as the \textit{Blandford-Znajek mechanism} after \citet{BZ77} who described it in terms of a stationary, axisymmetric, and force-free black hole magnetosphere.  They showed that such a magnetosphere can efficiently extract an enormous amount of energy, and ever since black hole energy extraction has been considered as a possible mechanism powering a range of astrophysical phenomena, from stellar mass black holes in gamma-ray bursts to supermassive black holes in active galactic nuclei.    

The Blandford-Znajek mechanism is not the only way to extract black hole rotational energy (\citet{LGATN2014} reviews others), nor is it strictly necessary for magnetic field lines to cross the horizon in order to extract black hole rotational energy (\citet{Komissarov2004} and references therein).  However the simple idea of a single magnetic field line connecting the horizon to distant observers can be seductive, as it is reminiscent of the more familiar and well understood case of a magnetic field line threading and torquing a rotating body.  As one might na\"{i}vely expect from such an analogy, \citet{BZ77} showed that for black hole energy to be extracted along a given magnetic field line that field line must rotate more slowly than the black hole.  In \citet{TNTT90} it was demonstrated that the condition of a relatively slowly rotating magnetic field line is actually the force-free limit of the requirement that a plasma inflow become super-Alfv\'{e}nic inside the ergosphere.  

Despite the defining role that the location of the inner Alfv\'{e}n surface plays in determining whether or not a given magnetosphere extracts black hole rotational energy, it is currently unknown how the shape and location of the Alfv\'{e}n surface might correspond to changes in the structure of a magnetosphere.  The differences (if any) between a magnetosphere with an Alfv\'{e}n surface near the horizon and a magnetosphere with an Alfv\'{e}n surface near the boundary of the ergosphere are largely uncertain.  This is due to the extreme intractability of the equations governing plasma flows in rotating black hole spacetimes, most especially the transfield equation that describes the force balance transverse to magnetic field lines (\citet{Nitta1991}).  Currently no exact solutions of magnetospheres with plasma flows satisfying the transfield equation are known, so there is no robust analytic method of studying any potential correspondences between Alfv\'{e}n surface location and magnetosphere structure.

In the force-free limit plasma inertial effects are ignored, the inner Alfv\'{e}n surface coincides with the inner light surface described by the field line angular velocity of the magnetosphere, and the transfield equation simplifies to a purely electromagnetic condition.  Due to those simplifications it is natural to consider the limit of force-free magnetospheres as a first step in attempting a study of the effects of Alfv\'{e}n surface location.  Unfortunately despite its relative simplicity the force-free limit of the transfield equation has very few analytic solutions, primarily limited to a handful of energy extracting magnetospheres found by perturbing in black hole spin (e.g. \citet{BZ77}, \citet{MG2004}, \citet{PanYu2015}) and a single class of fully exact but non energy extracting (and mostly unphysical) solutions (\citet{MenonDermer2007} and \citet{GrallaJacobson2014}).  Therefore even in the force-free limit there is also no robust analytic method of studying the effects of Alfv\'{e}n surface location.   

Current numerical solutions and simulations also lack significant utility, as the nature of the Alfv\'{e}n surface is ultimately an unknown function of the specific choice of boundary conditions and related initial assumptions that are made.  This means that the Alfv\'{e}n surface at best a result to be reported rather than an input to be explored, making it difficult to study how differing Alfv\'{e}n surfaces might correspond to changes in magnetosphere structure.     

The goal of this work was to study how Alfv\'{e}n surface location might correspond to the structure of energy extracting black hole magnetospheres, thereby beginning to fill a gap in our understanding of black hole energy extraction.  Due to the lack of useful analytic solutions it was necessary to calculate magnetospheres numerically.  For simplicity we focused on the limit of force-free magnetospheres.  We also focused on magnetospheres with a boundary condition in the equatorial plane compatible with a monopolar geometry; this avoids the confounding affects of the arbitrarily large forcings alternative boundary conditions can exert on the magnetosphere.  This boundary condition is also compatible with those used to calculate magnetospheres in \citet{CKP2013} and \citet{NC2014} as well as the perturbed monopole solutions originally calculated by \citet{BZ77}, allowing us to more easily compare our results with those works.     

To form an initial survey it was necessary to calculate a large number of magnetospheres for a wide range of potential black hole spins and field line angular velocities (ultimately $\sim500$ magnetospheres were found), so we developed numerical techniques that are efficient at calculating magnetospheres with similar structures.               

Those numerical techniques are described in Section \ref{Sec:Numerical}.  In Section \ref{Sec:Background} we provide some background on both ideal and force-free magnetospheres as well as the assumptions and equations we use.  In Section \ref{Sec:Results} we state some of our results, which we discuss in more depth in Section \ref{Sec:Discussion} before concluding.   
 
%%%%%%%%%%%%%%%%%%%%%%%%%%%%%%%%%%%%%%%%%%%%%%%%%%%%%%%%%%%%%%%%%%%%%%%%%%%%%%%%%%%%%%%%%%%%%%%%%%%%%%%%%%%%%%%%%%%%%%%%%%%%%%%%%%%%%%
%%%%%%%%%%%%%%%%%%%%%%%%%%%%%%%%%%%%%%%%%%%%%%%%%%%%%%%%%%%%%%%%%%%%%%%%%%%%%%%%%%%%%%%%%%%%%%%%%%%%%%%%%%%%%%%%%%%%%%%%%%%%%%%%%%%%%%
%%%%%%%%%%%%%%%%%%%%%%%%%%%%%%%%%%%%%%%%%%%%%%%%%%%%%%%%%%%%%%%%%%%%%%%%%%%%%%%%%%%%%%%%%%%%%%%%%%%%%%%%%%%%%%%%%%%%%%%%%%%%%%%%%%%%%%

\section{Background and Assumptions} \label{Sec:Background}

In the stationary and axisymmetric limit, force-free black hole magnetospheres are naturally described by three scalar fields: the toroidal vector potential $A_\phi (r, \theta)$, the field line angular velocity $\Omega_\text{F} (A_\phi)$, and the toroidal magnetic field $B_\phi (A_\phi)$.  In order to form a valid solution those scalar fields must satisfy a single differential equation, the force-free limit of the general relativistic Grad-Shafranov (or transfield) equation.  For black hole rotational energy to be extracted along a given field line that field line must rotate slower than the black hole; $0 < \Omega_\text{F} < \omega_\text{H}$, where $\omega_\text{H}$ is the angular velocity of the horizon.  In this section we provide a brief review of those results and the assumptions they rely on.  With the primary exception of $B_\phi$, which we define with opposite sign, the notation used is largely compatible with that of \citet{BZ77} and \citet{TNTT90} to facilitate comparison with those works; significant discrepancies will be noted.  Throughout we use covariant 4-vector notation; \citet{ThorneMacdonald1982} and \citet{Komissarov2004} provide reviews of and translations to the 3+1 formulations often used in other works.

%%%%%%%%%%%%%%%%%%%%%%%%%%%%%%%%%%%%%%%%%%%%%%%%%%%%%%%%%%%%%%%%%%%%%%%%%%%%%%%%%%%%%%%%%%%%%%%%%%%%%%%%%%%%%%%%%%%%%%%%%%%%%%%%%%%%%%
\subsection{Core Assumptions} \label{Sec:CoreAss}
%%%%%%%%%%%%%%%%%%%%%%%%%%%%%%%%%%%%%%%%%%%%%%%%%%%%%%%%%%%%%%%%%%%%%%%%%%%%%%%%%%%%%%%%%%%%%%%%%%%%%%%%%%%%%%%%%%%%%%%%%%%%%%%%%%%%%%

We begin by assuming a stationary, axisymmetric, uncharged, and isolated black hole such that the spacetime may be described by the Kerr metric in Boyer-Lindquist coordinates.  Using a $(+, -, -, -)$ metric signature and units where $c = G = 1$, this corresponds to the line element:
\begin{align}
ds^2 &= \left(1 - \frac{2mr}{\Sigma} \right) dt^2 + \frac{4mar \sin^2 \theta}{\Sigma} dt d\phi - \frac{\Sigma}{\Delta} dr^2 \nonumber \\
&- \Sigma d\theta^2 - \frac{A \sin^2 \theta}{\Sigma} d\phi^2,
\end{align}
where
\begin{align}
\Sigma &= r^2 + a^2 \cos^2 \theta ,\nonumber \\
\Delta &= r^2 - 2mr + a^2, 	\nonumber \\
A &= \left(r^2 + a^2 \right)^2 - \Delta a^2 \sin^2 \theta .
\end{align}

The black hole is then immersed in a perfectly conducting (ideal) plasma and electromagnetic fields that together lack sufficient energy density to modify the background spacetime.  The plasma and electromagnetic field configuration is also assumed to be stationary and axisymmetric, with an axis of symmetry corresponding to the rotational axis of the black hole.  The electromagnetic fields must be consistent with Maxwell's equations in a coordinate basis:
\begin{gather} 
\left(\sqrt{-g} F^{\alpha \beta} \right)_{, \beta} = -4 \pi \sqrt{-g} J^\alpha, \nonumber \\
F_{\alpha \beta, \gamma} + F_{\beta \gamma, \alpha} + F_{\gamma \alpha, \beta} = 0. \label{Eqn:MaxwellEqs}
\end{gather}  
Here $F^{\alpha \beta}$ is the field strength tensor, $J^\alpha$ is the current four vector, $g$ is the metric determinant, and a comma denotes a partial derivative.  We use Greek letters to denote spacetime indices $(\alpha = t, r, \theta, \phi$), lowercase Latin letters to denote spatial indices ($a = r, \theta, \phi$), and uppercase Latin letters to denote poloidal indices ($A = r, \theta$). The field strength tensor may be expressed in terms of an electromagnetic vector potential $A_\alpha$ as:
\begin{equation}
F_{\alpha \beta} = A_{\beta, \alpha} - A_{\alpha, \beta}.
\end{equation}  
In this form it is clear that the toroidal electric field vanishes ($F_{t \phi} = 0$) due to the assumptions of stationarity and axisymmetry, so ``toroidal field'' always references a toroidal magnetic field.  The stress energy tensor outside the horizon is taken to be a linear combination of plasma and electromagnetic effects:
\begin{align}
T^{\alpha \beta} &= T^{\alpha \beta}_\text{Plasma} + T^{\alpha \beta}_\text{EM} \nonumber \\
&= \left(\rho + p \right) u^\alpha u^\beta - p g^{\alpha \beta} \nonumber \\
&+\frac{1}{4 \pi} g^{\alpha \mu} F_{\mu \lambda} F^{\lambda \beta} + \frac{1}{16 \pi} g^{\alpha \beta} F_{\mu \lambda} F^{\mu \lambda},
\end{align}
where $\rho$ and $p$ reference the proper energy density and pressure of the plasma, and $u^\alpha$ its four velocity.

We finally assume that there is no external forcing (energy and momentum is conserved) over any region of interest such that the divergence of the stress energy tensor vanishes everywhere:
\begin{equation}
T^{\alpha \beta}{}{}_{; \beta} = 0.
\end{equation}
Everything that follows rests upon the above assumptions and conditions.    

%%%%%%%%%%%%%%%%%%%%%%%%%%%%%%%%%%%%%%%%%%%%%%%%%%%%%%%%%%%%%%%%%%%%%%%%%%%%%%%%%%%%%%%%%%%%%%%%%%%%%%%%%%%%%%%%%%%%%%%%%%%%%%%%%%%%%%
\subsection{Field Aligned Conserved Quantities}
%%%%%%%%%%%%%%%%%%%%%%%%%%%%%%%%%%%%%%%%%%%%%%%%%%%%%%%%%%%%%%%%%%%%%%%%%%%%%%%%%%%%%%%%%%%%%%%%%%%%%%%%%%%%%%%%%%%%%%%%%%%%%%%%%%%%%%

The three scalar fields that describe a force-free magnetosphere (vector potential $A_\phi$, field line angular velocity $\Omega_\text{F}$, and toroidal magnetic field $B_\phi$) are field-aligned conserved quantities in the sense that for a scalar field $\Psi$ we have:
\begin{equation}
B^\alpha \Psi_{, \alpha} = 0.
\end{equation}
Here $B^\alpha$ is the magnetic field measured by a distant observer, which we define as:
\begin{equation} \label{MagneticFieldDef}
B^\alpha = \frac{1}{2 \sqrt{-g}} \epsilon^{\alpha \beta \mu \nu} k_\beta F_{\mu \nu},
\end{equation} 
where $k_\beta$ is the Killing vector associated with the stationarity of the spacetime.  

The toroidal vector potential $A_\phi$ directly describes the poloidal magnetic field, making it a useful flux function to trace poloidal magnetic field lines.  The conservation of $A_\phi$ rests upon stationarity and axisymmetry, so $A_\phi$ is also a valid flux function in non force-free magnetospheres.   

The conservation of field line angular velocity $\Omega_\text{F}$ is the general relativistic extension of Ferraro's law of isorotation. The field line angular velocity is defined in terms of the field strength tensor as:
\begin{gather}
F_{r \phi} \Omega_\text{F} = F_{tr}, \nonumber \\
F_{\theta \phi} \Omega_\text{F} = F_{t \theta}.
\end{gather}
This states that all electric fields are the result of the rotation of a purely magnetic configuration with angular velocity $\Omega_\text{F}$ (referenced to zero angular momentum frames).  The rigid rotation of individual magnetic field lines through the conservation of $\Omega_\text{F}$ is a consequence of stationarity, axisymmetry, and a perfectly conducting plasma, so it is also valid in non force-free magnetospheres.  

The toroidal field $B_\phi$ is defined through the definition of the magnetic field $B^\alpha$ in Equation \ref{MagneticFieldDef} as $B_\phi = \sqrt{-g} F^{\theta r}$.  This has opposite sign to the definition in \citet{BZ77}, so for the remainder of this work the toroidal field will be referred to as $\sqrt{-g} F^{\theta r}$ to avoid confusion.  The conservation of the toroidal field may be understood by considering Poynting fluxes through the magnetosphere.  The only electric field is the result of rotating a purely magnetic configuration, and is therefore both entirely poloidal and perpendicular to the poloidal magnetic field.   Therefore all poloidal Poynting fluxes are weighted only by the toroidal magnetic field and aligned with the poloidal magnetic field.  As such the toroidal field is naturally interpreted as a measure of a conserved flux of energy $E$ and angular momentum $L$ per unit field line:
\begin{align} \label{EandLEquation}
E &= \frac{1}{4 \pi} \sqrt{-g} F^{\theta r} \Omega_\text{F}, \nonumber \\
L &= \frac{1}{4 \pi} \sqrt{-g} F^{\theta r}.
\end{align}
Note that these definitions differ from those in \citet{TNTT90}; in that work they are scaled by a conserved particle flux $\eta$ as 
$E \rightarrow \eta E$ and $L \rightarrow \eta L$.  The conservation of the toroidal field along magnetic field lines is a consequence of stationarity, axisymmetry, and a perfectly conducting force-free plasma; in non force-free magnetospheres $E$ and $L$ include plasma parameters and $\sqrt{-g}F^{\theta r}$ is not conserved.

Varied derivations and discussions of these and other field-aligned conserved quantities may be found in \citet{BZ77}, \citet{BekensteinOron1978}, \citet{Camenzind1986a}, and \citet{TNTT90}.

%%%%%%%%%%%%%%%%%%%%%%%%%%%%%%%%%%%%%%%%%%%%%%%%%%%%%%%%%%%%%%%%%%%%%%%%%%%%%%%%%%%%%%%%%%%%%%%%%%%%%%%%%%%%%%%%%%%%%%%%%%%%%%%%%%%%%%
\subsection{Critical Points and Energy Extraction}
%%%%%%%%%%%%%%%%%%%%%%%%%%%%%%%%%%%%%%%%%%%%%%%%%%%%%%%%%%%%%%%%%%%%%%%%%%%%%%%%%%%%%%%%%%%%%%%%%%%%%%%%%%%%%%%%%%%%%%%%%%%%%%%%%%%%%%

In this section we explore key points along magnetic field lines and examine the conditions required for a magnetosphere to extract black hole rotational energy.  We begin by making two assumptions that are unnecessary in general but greatly simplify the discussion.  First, we assume a single magnetic field line that extends directly from the horizon to a distant observer.  Second, we insist that this field line be ``simple'' in the sense that the distance from the horizon decreases monotonically as one travels along the field line towards the black hole; a useful mental image is a straight line connecting the horizon with distant regions.  

Along any such field line there are seven key points, the first being the separation point.  Inside the separation point gravitational forces dominate and draw plasma inwards to the horizon; outside the separation point centrifugal forces in the rotating frame of the magnetic field line dominate and accelerate plasma away from the black hole.  The remaining six points are the ingoing and outgoing slow magnetosonic, Alfv\'{e}n, and fast magnetosonic points through which the plasma accelerates as it travels from the separation point to the horizon or distant regions.  In the force-free limit the ingoing slow magnetosonic point coincides with the separation point, the ingoing Alfv\'{e}n point coincides with the inner light surface, and the ingoing fast magnetosonic point coincides with the horizon.  

The inner and outer light surfaces are defined by the inner and outer surfaces exterior to the horizon where $\alpha = 0$ ($\alpha$ is defined in Equation \ref{alphaEqn} in the next section).  The outer light surface is analagous to a pulsar light cylinder, where rigid rotation would cause a particle stuck to a magnetic field line to rotate at the speed of light.  The particle would also rotate at the speed of light at the inner light surface as a consequence of gravitational time dilation, and it would be subluminal between the light surfaces.     

A field line extracts black hole rotational energy when its ingoing Alfv\'{e}n point is located inside the ergoregion.  In the force-free limit this reduces to the requirement that the magnetic field line rotate more slowly than the black hole:
\begin{equation}
0 < \Omega_\text{F} < \omega_\text{H}.
\end{equation}
Here $\omega_{H} = a/2 m r_+$ is the angular velocity of a zero angular momentum observer on the horizon.  
In general there exist regularity conditions at each of the slow magnetosonic, Alfv\'{e}n, and fast magnetosonic points.  In the force-free limit the most useful of these is that of the ingoing fast magnetosonic point, which reduces to a condition on the fields at the horizon:

\begin{equation} \label{ZnajekEqn}
\sqrt{-g} F^{\theta r} = \pm \frac{\left(r_+^2 + a^2 \right) \left(\Omega_\text{F} - \omega_\text{H} \right) \sin \theta}{\Sigma} F_{\theta \phi}.
\end{equation}

The equations in use are insensitive to the sign of the toroidal field, but if an ingoing observer on the horizon is to measure finite electromagnetic fields \citet{Znajek1977} showed that the negative branch must be chosen.  This ``Znajek condition'' does not place any additional restrictions on the fields.  It was already present in our core assumptions, most importantly the conservation of energy and momentum in the vanishing divergence of the stress energy tensor.  However it can be a simple check to ensure that a solution is valid on the horizon. 

Further discussion of the key points along a magnetic field line may be found in \citet{TNTT90} and \citet{Takahashi2002}.  

%%%%%%%%%%%%%%%%%%%%%%%%%%%%%%%%%%%%%%%%%%%%%%%%%%%%%%%%%%%%%%%%%%%%%%%%%%%%%%%%%%%%%%%%%%%%%%%%%%%%%%%%%%%%%%%%%%%%%%%%%%%%%%%%%%%%%%
\subsection{The Force-Free Transfield Equation} \label{Sec:FFTrans}
%%%%%%%%%%%%%%%%%%%%%%%%%%%%%%%%%%%%%%%%%%%%%%%%%%%%%%%%%%%%%%%%%%%%%%%%%%%%%%%%%%%%%%%%%%%%%%%%%%%%%%%%%%%%%%%%%%%%%%%%%%%%%%%%%%%%%%

In our core assumptions we demanded that there be no external forcing on the magnetosphere, as stated in the requirement that $T^{\alpha \beta}{}{}_{; \beta} = 0$.  If the poloidal components of this requirement are satisfied then the temporal and toroidal components are automatically satisfied.  We therefore focus on the two poloidal components, which may be reduced to a single differential equation; this equation is often referred to as the ``transfield equation'' as it encapsulates the force balance transverse to poloidal magnetic field lines.  In the force-free limit the transfield equation is given by (cf. Equation 3.14 of \citet{BZ77}):    
\begin{align} \label{TransfieldEqn}
0 &= \frac{1}{2} \frac{\Sigma}{\Delta \sin \theta} \frac{d}{d A_\phi} \left(\sqrt{-g} F^{\theta r} \right)^2 \nonumber \\
&+ \frac{1}{\sin \theta} \left[\alpha F_{r \phi} \right]_{, r} + \frac{1}{\Delta} \left[ \frac{\alpha}{\sin \theta} F_{\theta \phi}\right]_{, \theta} \nonumber \\
&- \frac{G_\phi}{\sin \theta} \left(F_{r \phi} \Omega_{\text{F}, r} + \frac{1}{\Delta} F_{\theta \phi} \Omega_{\text{F}, \theta} \right),
\end{align}
where 
\begin{align} \label{alphaEqn}
\alpha &= g_{tt} + 2 g_{t \phi} \Omega_\text{F} + g_{\phi \phi} \Omega_\text{F}^2, \nonumber \\
G_\phi &= g_{t \phi} + g_{\phi \phi} \Omega_\text{F}. 
\end{align}
The function $\alpha^{-1/2}$ may be interpreted as the ``gravitational Lorentz factor'' of a particle stuck to a magnetic field line, describing the effects of both rotation and gravitational time dilation, while $G_\phi$ may be interpreted as a measure of that particle's rotational velocity with respect to the spacetime.           

Solving the transfield equation is the core difficulty of finding black hole magnetospheres.  In the stationary and axisymmetric limit of a rotating black hole, only one exact class of solutions is currently known (first published by \citet{MenonDermer2007} and discussed in a more general framework by \citet{GrallaJacobson2014}):
\begin{align} \label{ExactSolnEqn}
A_\phi &= B_0 h(\theta), \nonumber \\
\Omega_\text{F} &= \frac{1}{a \sin^2 \theta}, \nonumber \\
\sqrt{-g} F^{\theta r} &= B_0 \frac{1}{a} g(\theta).
\end{align} 
Here $B_0$ is a scalar measure of the strength of the poloidal magnetic field; $h$ and $g$ are arbitrary functions of $\theta$ that must satisfy (prime denoting a derivative with respect to $\theta$):
\begin{equation} \label{ExactSolnCondEqn}
h' = -g \sin \theta.
\end{equation}
Solutions of this type generically contain an unphysical divergence in $\Omega_\text{F}$ in both the $\theta \rightarrow 0$ and low black hole spin $a \rightarrow 0$ limits, and are therefore mostly mathematical curiosities.  They also do not extract rotational energy from the black hole along any field lines, further limiting their interest.  However we still find them to be useful tools in determining the potential precision of a given numerical grid; see Sections \ref{Sec:AnalyticComp} and \ref{Sec:NumErrRes}.      

In the limit of a non-rotating ($a = 0$) Schwarzschild black hole, there does exist a more reasonable variety of exact rotating ``monopolar'' solutions, given by:
\begin{align} \label{SchwMonEqn}
A_\phi &= B_0 \cos \theta, \nonumber \\
\Omega_\text{F} &= \Omega_0, \nonumber \\
\sqrt{-g} F^{\theta r} &= B_0 \Omega_0 \sin^2 \theta.
\end{align} 
Such solutions do not extract rotational energy from the black hole (it has none to extract) but nontheless serve as a convenient starting point for exploring black hole energy extraction.  These solutions are sometimes stated as ``split-monopole'' solutions by changing the sign of $B_0$ below the equatorial plane and appealing to an equatorial current sheet.  For simplicity we will always refer to them as ``monopolar'' solutions, as regardless of interpretation there is still reflection symmetry across the equatorial plane.  However it should be noted that ``monopolar'' references only the poloidal magnetic field; there is often a significant toroidal field, as can be seen in the $\Omega_0 \neq 0$ solutions.  Setting $\Omega_0 = 0$ for a static monopole and applying that solution to a slowly rotating black hole by perturbing in black hole spin (as in \citet{BZ77} and more recently \citet{PanYu2015}) yields:
\begin{align} \label{PertMonEqn}
A_\phi &= B_0 \cos \theta - \left\{ \frac{a^2}{m^2} B_0 R\left(r\right) \cos \theta \sin^2 \theta \right\}, \nonumber \\
\Omega_\text{F} &= \frac{a}{8m^2} + \left\{\frac{a^3}{32m^4} \left[1 + \frac{1}{2} \left(1 - 2R_2 \right) \sin^2 \theta \right] \right\}, \nonumber \\
\sqrt{-g} F^{\theta r} &= -B_0 \frac{a}{8 m^2} \sin^2 \theta \nonumber \\
&-  \left\{ \frac{ B_0 a^3 \sin^2 \theta}{32 m^4} \left[1 + \frac{1}{2} \left(1 - 4 R_2 \right) \sin^2 \theta \right. \right. \nonumber \\
&+ \left. \vphantom{\frac{ B_0 a^3 \sin^2 \theta}{32 m^4}} \left. \vphantom{\frac{1}{2}} 8 R\left(r \right) \cos^2 \theta \right] \right\},
\end{align}
where
\begin{align}
R\left( r \right) &= \frac{1}{m^2} \left[\frac{m^2 + 3mr - 6r^2}{12} \ln \left(\frac{r}{2m} \right) + \frac{11 m^2}{72} \right. \nonumber \\
&+ \left. \vphantom{\frac{m^2 + 3mr - 6r^2}{12} \ln \left(\frac{r}{2m} \right)} \frac{m^3}{3r} + \frac{mr}{2} - \frac{r^2}{2} \right] \nonumber \\
&+ \left(\frac{2r^3 - 3mr^2}{8m^3} \right) \left[\text{Li}_2 \left(\frac{2m}{r}\right) \right. \nonumber \\
&- \left. \vphantom{\text{Li}_2 \left(\frac{2m}{r}\right)} \ln \left(1 - \frac{2m}{r} \right) \ln \left(\frac{r}{2m} \right) \right], \nonumber \\
R_2 &= \frac{1}{72} \left( 6 \pi^2 -49 \right), 
\end{align}  
and $\text{Li}_2$ is the dilogarithm, defined as:
\begin{equation}
\text{Li}_2 (x) = \int_x^0 \frac{1}{t} \ln \left(1 -t \right) dt.
\end{equation}
The higher order corrections by \citet{PanYu2015} (and \citet{MG2004}) are shown in curly brackets.  It is possible to extend the above to include $\mathcal{O}(a^5)$ corrections, but the solutions achieve a level of complication far surpassing the point of reason.  There are two things to note from the above expressions; first, and most obviously, is that attempting to find perturbed solutions is not a trivial undertaking.  Perturbing one of the simplest possible magnetic field configurations is effectively impossible past $\mathcal{O}(a^2)$ corrections to $A_\phi$.  The second thing to note that the leading order correction to $\Omega_0 = 0$ is $\Omega_\text{F} = 0.5 \omega_\text{H}$ (to first order in $a$).  This has partly contributed to the ``rule of thumb'' that energy extracting black hole magnetospheres should rotate at one half the rate of the black hole ($\Omega_\text{F} \approx 0.5 \omega_\text{H}$ also extracts the maximum amount of black hole energy for low spin).      

Exploring the $\Omega_0 \neq 0$ solutions was one of the initial questions that prompted our work, as they naturally encompass a full range of inner Alfv\'{e}n surfaces and corresponding magnetospheres.  Unfortunately analytically perturbing around such solutions is even more intractable than the already non-trivial $\Omega_0 = 0$ case, so we developed numerical techniques to solve for the structure of black hole magnetospheres.  Discussion of those techniques is the topic of Section \ref{Sec:Numerical}.  

%%%%%%%%%%%%%%%%%%%%%%%%%%%%%%%%%%%%%%%%%%%%%%%%%%%%%%%%%%%%%%%%%%%%%%%%%%%%%%%%%%%%%%%%%%%%%%%%%%%%%%%%%%%%%%%%%%%%%%%%%%%%%%%%%%%%%%
\subsection{Additional Assumptions} \label{Sec:ModelAss}
%%%%%%%%%%%%%%%%%%%%%%%%%%%%%%%%%%%%%%%%%%%%%%%%%%%%%%%%%%%%%%%%%%%%%%%%%%%%%%%%%%%%%%%%%%%%%%%%%%%%%%%%%%%%%%%%%%%%%%%%%%%%%%%%%%%%%%

In addition to the core assumptions of Section \ref{Sec:CoreAss}, we make two additional simplifying assumptions guided by our goal of studying correspondences between inner Alfv\'{e}n surface location and magnetosphere structure.  The first is the assumption of a monopolar geometry, in the sense that the poloidal magnetic field line threading the horizon on the equator must remain on the equator.  This assumption is enforced via a boundary condition on the flux function $A_\phi$; we require that $A_\phi$ be constant in the equatorial plane.  In making this assumption we were guided by a desire to extend the Schwarzschild monopole solutions of Equation  \ref{SchwMonEqn} to rotating spacetimes and to avoid assuming a specific model for nearby accreting matter.  Without the presence of such matter to restrict magnetic field lines, a configuration with varying vector potential in the equatorial plane could violate the condition that there can be no closed horizon loops in stationary force-free magnetospheres (see \citet{MacDonaldThorne1982} and additional discussion in \citet{GrallaJacobson2014}).  Additonally, boundary conditions encoding an assumed model of matter in the equatorial plane or elsewhere have the potential to exert arbitrarily large forcings on the magnetosphere, confounding attempts to explore basic correlations between Alfv\'{e}n surface location and magnetosphere structure.  The interested reader may see \citet{Uzdensky2005} for some complications induced by horizon-disk connections and \citet{Fendt1997} for disk-jet connections.

The second assumption we made is that of uniform field line angular velocity, as that is the simplest possible method of classifying inner light surfaces (inner Alfv\'{e}n surfaces in the force-free limit) and allowed us to know exactly where the inner light surface would be prior to calculating the structure of the field lines that pass through it.  While it is convenient, this assumption restricts us to a region outside the horizon that is bounded by the outer light surface.  This is due to the fact that our dissipative numerical techniques cannot generically find magnetospheres that pass smoothly through both inner and outer light surfaces under the assumption of uniform field line angular velocity, even if such solutions exist (we discuss smoothness at light surfaces in Section \ref{Sec:Kinks}).  Ultimately we do not believe this to be nearly as limiting an assumption as it might initially appear, but it is still the most awkward assumption that we have made.  As such we return to it in Section \ref{Sec:InitialAss} with more detailed discussion.

%%%%%%%%%%%%%%%%%%%%%%%%%%%%%%%%%%%%%%%%%%%%%%%%%%%%%%%%%%%%%%%%%%%%%%%%%%%%%%%%%%%%%%%%%%%%%%%%%%%%%%%%%%%%%%%%%%%%%%%%%%%%%%%%%%%%%%
%%%%%%%%%%%%%%%%%%%%%%%%%%%%%%%%%%%%%%%%%%%%%%%%%%%%%%%%%%%%%%%%%%%%%%%%%%%%%%%%%%%%%%%%%%%%%%%%%%%%%%%%%%%%%%%%%%%%%%%%%%%%%%%%%%%%%%
%%%%%%%%%%%%%%%%%%%%%%%%%%%%%%%%%%%%%%%%%%%%%%%%%%%%%%%%%%%%%%%%%%%%%%%%%%%%%%%%%%%%%%%%%%%%%%%%%%%%%%%%%%%%%%%%%%%%%%%%%%%%%%%%%%%%%%

\section{Numerical Techniques} \label{Sec:Numerical}

In order to numerically solve for a large number of black hole magnetospheres, we have extended the Newtonian magnetofrictional method developed by \citet{YSA1986} to the general relativistic regime.  The magnetofrictional method makes an initial guess as to the structure of a force-free magnetosphere and then gradually relaxes the fields towards a valid force-free state.  This makes it a useful tool for exploring our parameter space; once a solution for a given black hole spin and field line angular velocity is known adjacent solutions for slightly different values are readily found.  In this section we discuss the underlying theory of the magnetofrictional method as well as the computational specifics of our implementation.

%%%%%%%%%%%%%%%%%%%%%%%%%%%%%%%%%%%%%%%%%%%%%%%%%%%%%%%%%%%%%%%%%%%%%%%%%%%%%%%%%%%%%%%%%%%%%%%%%%%%%%%%%%%%%%%%%%%%%%%%%%%%%%%%%%%%%%
\subsection{The Magnetofrictional Method}
%%%%%%%%%%%%%%%%%%%%%%%%%%%%%%%%%%%%%%%%%%%%%%%%%%%%%%%%%%%%%%%%%%%%%%%%%%%%%%%%%%%%%%%%%%%%%%%%%%%%%%%%%%%%%%%%%%%%%%%%%%%%%%%%%%%%%%

Once initial guesses have been made for a vector potential $A_\phi$, field line angular velocity $\Omega_\text{F}$, and toroidal field $\sqrt{-g} F^{\theta r}$, they may be inserted into the transfield equation to see if those guesses form a valid solution.  This is equivalent to determining if the poloidal components of the divergence of the stress energy tensor vanish: 
\begin{equation}
T^{A \beta}{}{}_{; \beta} = X^A.
\end{equation}
If the initial guesses formed a valid and self consistent force-free solution then $X^A$ would vanish, but in general $X^A \neq 0$ and there is an excess momentum flux that may be interpreted as an external forcing.  The core idea of the magnetofrictional method is to take that external forcing and convert it into the coordinate velocity of a fictitious plasma via friction.  In this way a physically invalid (or at least inconsistent) initial guess for a force-free magnetosphere may be converted into a valid non force-free magnetosphere, and the fields may be evolved towards a force-free state by applying the equations governing inertial plasma flows.  The first step in this process is to evaluate the excess momentum flux $X^A$, which is found as a modification to the left hand side of the transfield equation (Equation \ref{TransfieldEqn}):   
\begin{equation}
- 4 \pi \Sigma \sin \theta \frac{X_A}{A_{\phi, A}} = \frac{1}{2} \frac{\Sigma}{\Delta \sin \theta} \frac{d}{d A_\phi} \left(\sqrt{-g} F^{\theta r} \right)^2  + \ldots
\end{equation}
Once the excess momentum flux has been determined, it may be converted into the coordinate velocity of a fictitious plasma $v^A$ ($v^r = \partial r/\partial t$, $v^\theta = \partial \theta / \partial t$) via friction:  
\begin{equation} \label{Eqn:NuEqn}
X^A = \frac{1}{\nu} v^A.
\end{equation}     
Here the coefficient $\nu$ measures the strength of the friction.  We then assume that the fictitious plasma is a perfect conductor such that there is no electric field in its rest frame:
\begin{equation}
F_{\alpha \beta} u^\beta = 0.
\end{equation}
Here $u^\alpha$ is the four velocity of the plasma, which we define in terms of its coordinate velocity as $v^c \equiv u^c / u^t$.  The above may then be used to find:
\begin{equation}
F_{tc} = F_{cb} v^b.
\end{equation} 
If we insert this relationship into Maxwell's homogeneous equations (Equation \ref{Eqn:MaxwellEqs}) we find the relativistic analog of the ideal induction equation $\partial_t \mathbf{B} = \mathbf{\nabla} \mathbf{\times} (\mathbf{v} \mathbf{\times} \mathbf{B})$:     
\begin{equation}
F_{ab, t} = \left(F_{bc} v^c \right)_{, a} - \left(F_{ac} v^c \right)_{, b}.
\end{equation}
Inserting either $F_{\theta \phi}$ or $F_{r \phi}$ into the above then shows that the time rate of change of the vector potential $A_\phi$ is given by:
\begin{equation} \label{Eqn:AdvecEqn}
A_{\phi, t} = -v^A A_{\phi, A}.
\end{equation}
In Appendix \ref{App:ConvProof} we prove that application of this equation inevitably leads to a force-free configuration and in Appendix \ref{App:MFExpand} we expand it in Boyer-Lindquist coordinates.  Although the coordinate velocity $v^A$ is a complicated function of the metric and electromagnetic fields, at a high level the magnetofrictional method reduces to the straightforward evolution of an advection equation.  A variety of numerical techniques for evolving such equations exist.  For simplicity we have implemented an upwind differencing scheme similar to that described in \citet{HSW1984}.  More sophisticated techniques might  converge on a valid solution more rapidly, but we found upwind differencing to be both very numerically stable and fast enough for our purposes.  

We also note that the relaxation scheme implemented by \citet{Uzdensky2005} is similar to the method described above, in that it uses $A_{\phi, t} \sim v^A$; the primary difference is in the weighting of $v^A$.

%%%%%%%%%%%%%%%%%%%%%%%%%%%%%%%%%%%%%%%%%%%%%%%%%%%%%%%%%%%%%%%%%%%%%%%%%%%%%%%%%%%%%%%%%%%%%%%%%%%%%%%%%%%%%%%%%%%%%%%%%%%%%%%%%%%%%%
\subsection{Fixing Kinks} \label{Sec:Kinks}
%%%%%%%%%%%%%%%%%%%%%%%%%%%%%%%%%%%%%%%%%%%%%%%%%%%%%%%%%%%%%%%%%%%%%%%%%%%%%%%%%%%%%%%%%%%%%%%%%%%%%%%%%%%%%%%%%%%%%%%%%%%%%%%%%%%%%%

The magnetofrictional method described above is limited by the development of kinks in the magnetic field, which we now discuss.  The transfield equation (Equation \ref{TransfieldEqn}) changes character at the inner and outer light surfaces where the gravitational Lorentz factor $\alpha$ changes sign.  This change in character is transmitted to the coordinate velocity $v^A$ of the fictitious plasma, which goes as:
\begin{equation}
v^A \sim \alpha \left( A_{\phi, rr} + A_{\phi, \theta \theta} \right).
\end{equation}
On the inner and outer light surfaces $\alpha$ vanishes; it is positive between them and negative outside.  To avoid having our numerical scheme become anti-diffusive and unstable, outside the light surfaces we introduce an overall multiplicative factor of $-1$ to maintain stability.  This factor may be thought of as a correction for the fact that the fictitious plasma introduced by the magnetofrictional method is superluminal outside the light surfaces, and does not affect convergence. 

However the change in sign of $\alpha$ does split the computational domain into three regions, and in general those regions do not join smoothly.  In other words for an initial guess of field line angular velocity $\Omega_\text{F}$ and toroidal field $\sqrt{-g} F^{\theta r}$, the solutions that the magnetofrictional method finds for $A_\phi$ will in general not match across light surfaces.  Treating $A_\phi$ as a flux function, this leads to discontinuous ``kinks'' in the magnetic field at the inner and outer light surfaces.  As there are two light surfaces, both the field line angular velocity and the toroidal field could be evolved simultaneously in order to diminish both kinks, as in \citet{CKP2013} and \citet{NC2014}.  

However, as stated in Section \ref{Sec:ModelAss}, for simplicity we have chosen to use a fixed and uniform field line angular velocity so we do not have the freedom required to generically diminish both kinks using the inherently dissipative magnetofrictional method.  We therefore divide the computational domain into two regions, one inside the inner light surface and one between the light surfaces, and only diminish the kink at the inner light surface by evolving the toroidal field.  The limitations and advantages of this approach are discussed in Section \ref{Sec:InitialAss}.  

In order to measure the kink at the inner light surface, the magnetofrictional method is applied separately to the regions on either side until an empirical ``zero level'' in the velocity $v^A$ is reached.  A vanishing plasma coordinate velocity corresponds to $X^A = 0$ and a valid force-free configuration, so this procedure simply finds a ``close enough'' solution in both regions.  The two solutions are then compared across the inner light surface and the magnitude of any kink is measured in terms of $\Delta A_\phi$.  This measurement is then used to modify the toroidal field in a method similar to that developed by \citet{Contopoulos1999} for light cylinders in pulsar magnetospheres:
\begin{equation} \label{Eqn:KinkFix}
\frac{d}{d A_\phi} \left(\sqrt{-g} F^{\theta r}\right)^2_\text{New} = \left. \frac{d}{d A_\phi} \left(\sqrt{-g} F^{\theta r}\right)^2 \right|_{A_{\phi-}} - \gamma \Delta A_\phi.
\end{equation} 
Here $\gamma$ is an empirical constant, usually much less than one, and $A_{\phi-}$ denotes the value of $A_\phi$ just inside the inner light surface.  After the above equation has been used to make a small correction to the toroidal field, the magnetofrictional method is again applied to the regions on either side of the inner light surface for a fixed number of time steps, and then (if the set ``zero level'' in the plasma coordinate velocity is still achieved) the kink is re-evaluated and toroidal field re-adjusted.  The cycle repeats until matching solutions are found.

Finding the correct ``zero level'', value for $\gamma$, and number of time steps to apply the magnetofrictional method between modifications of the toroidal field is akin to critically damping a simple harmonic oscillator.  Poor choices can lead to either an underdamped, wildly oscillating kink or an overdamped kink that requires an excessive amount of computational time to diminish.  We have found that a ``zero level'' of $10^{-4}$, a factor of $\gamma = 0.0005$, and $3000$ time steps between toroidal field changes is typically a safe starting point, but optimal values can vary significantly depending upon everything from grid resolution to the current magnitude and nature of the kink and how good the initial guess for the toroidal field is.      

%%%%%%%%%%%%%%%%%%%%%%%%%%%%%%%%%%%%%%%%%%%%%%%%%%%%%%%%%%%%%%%%%%%%%%%%%%%%%%%%%%%%%%%%%%%%%%%%%%%%%%%%%%%%%%%%%%%%%%%%%%%%%%%%%%%%%%
\subsection{Measuring Force-Freeness} \label{Sec:MeasuringFF}
%%%%%%%%%%%%%%%%%%%%%%%%%%%%%%%%%%%%%%%%%%%%%%%%%%%%%%%%%%%%%%%%%%%%%%%%%%%%%%%%%%%%%%%%%%%%%%%%%%%%%%%%%%%%%%%%%%%%%%%%%%%%%%%%%%%%%%

In principle the magnetofrictional method could be applied to the limits of numerical precision.  However even if this were practical the approximations and assumptions we have made do not justify that level of precision.  Instead we evolve the fields until a ``good enough'' level is reached, which we now describe.  

A force-free solution is a solution with a vanishing Lorentz force, $F_{\alpha \beta} J^\beta = 0$.  Therefore it is reasonable to take a ratio of the Lorentz force to the magnitude of electric current and field strength tensor in order to measure how force-free a solution is, an approach taken in \citet{MG2004}:
\begin{equation}
\xi = \left|\frac{F^{\mu \nu} J_\nu F_{\mu k} J^k}{J_\mu J^\mu F_{\kappa \lambda} F^{\kappa \lambda}} \right|.
\end{equation}    
If $\xi \ll 1$ the configuration can be thought of as being force-free.  Unfortunately such a technique is of limited utility in our case; many of our magnetospheres contain regions of vanishing or very small current $J_\alpha J^\alpha \ll 1$, leading to divergences in $\xi$.  

In force-free magnetospheres with regions of vanishing current, one alternative to $\xi$ is to measure the ratio of the Lorentz force to the forces of magnetic pressure and tension, as in \citet{MSDW2014}:
\begin{equation}
\zeta = \frac{\left|\mathbf{F}_\text{L} \right|}{\left|\mathbf{F}_\text{mp} \right| + \left| \mathbf{F}_\text{mt} \right|}.
\end{equation} 
As with $\xi$,  if $\zeta \ll 1$ then the configuration may be taken to be force free.  Unfortunately some of our magnetospheres contain very strong monopolar components for which the forces of magnetic pressure and tension vanish separately, again leading to divergences.  We have therefore developed a more mathematical and less physically motivated measure of force-freeness.  We begin by noting that the transfield equation may be broken up into seven terms:
\begin{equation}
T_A{}^\beta{}_{; \beta} = -F_{A \beta} J^\beta = A_{\phi, A} \sum_{i=1}^7 D_i.
\end{equation}
The separation of the transfield equation into $D_i$ components is done somewhat arbitrarily by grouping factors of $A_{\phi}$, field line angular velocity, and the toroidal field; $D_1 \sim \sqrt{-g} F^{\theta r}$, $D_3 \sim A_{\phi, rr}$, etc.  The exact separations we use are expanded in Appendix \ref{PercentAppendix}.  Mathematically the above puts us in the position of having an expression of the form $\sum D_i = \delta$ and requiring that $\delta$ be ``small''.   As a measure of force-freeness, we therefore take the absolute maximum of the seven $D_i$ terms and require that $\delta$ be a factor of $\epsilon$ smaller than it:  
\begin{equation} \label{PercentErrorEqn}
\left|\delta\right| < \epsilon \cdot \text{Max} \left( \left|D_i \right| \right).
\end{equation}
While such a method is disadvantaged by not having an obvious physical interpretation, it does not diverge when applied to magnetospheres with regions of vanishing current or with magnetic fields containing strong monopolar components.  For this work we have selected $\epsilon = 1\%$ as a level of sufficient force-freeness.  In practice our numerical techniques result in only a few segments of the inner light surface approaching $\epsilon = 1\%$; over the rest of the domain $\epsilon$ is significantly smaller.   

%%%%%%%%%%%%%%%%%%%%%%%%%%%%%%%%%%%%%%%%%%%%%%%%%%%%%%%%%%%%%%%%%%%%%%%%%%%%%%%%%%%%%%%%%%%%%%%%%%%%%%%%%%%%%%%%%%%%%%%%%%%%%%%%%%%%%%
\subsection{Analytic Comparisons} \label{Sec:AnalyticComp}
%%%%%%%%%%%%%%%%%%%%%%%%%%%%%%%%%%%%%%%%%%%%%%%%%%%%%%%%%%%%%%%%%%%%%%%%%%%%%%%%%%%%%%%%%%%%%%%%%%%%%%%%%%%%%%%%%%%%%%%%%%%%%%%%%%%%%%

In Section \ref{Sec:FFTrans} we discussed some analytic solutions to the force-free transfield equation.  Here we select four specific solutions that are useful for making comparisons with our numerical solutions; in Section \ref{Sec:NumErrRes} we compare them to a numerical solution with $\Omega_\text{F} = 0.5 \omega_\text{H}$ and $a = 0.3 m$ in order to examine the consequences of the $\epsilon = 1\%$ error level discussed in the previous section.

The first analytic solution that provides a useful comparison is the exact Schwarzschild monopole described by Equation \ref{SchwMonEqn}.  The field line angular velocity $\Omega_0$ of this solution may be set to correspond to the field line angular velocity of a given numerical solution.  The monopole weight $B_0$ is set to $B_0=1$ in order to match the numerical solution's change in $A_\phi$ between the axis and the equator (discussed in the next section).  This solution provides a comparison with all field line angular velocities but is most useful for very small spins.
 
The second useful solution is the first order perturbed monopole solution (Equation \ref{PertMonEqn} without the terms in curly brackets).  This solution is very similar to the Schwarzschild monopole with $\Omega_0 = 0.5 \omega_\text{H}$, but differs slightly due to the fact that its field line angular velocity is only equivalent to $0.5 \omega_\text{H}$ to first order in spin parameter $a$.  As with the Schwarzschild monopole solution we set $B_0 = 1$. The field line angular velocity of this solution is fixed, so it is most useful for comparison with numerical solutions that have $\Omega_\text{F} = 0.5 \omega_\text{H}$ and small black hole spins.

The third solution is the third order perturbed monopole solution described by Equation \ref{PertMonEqn}.  As with the first order solution we set $B_0 = 1$.  This solution is most useful for comparison with numerical solutions of $\Omega_\text{F} = 0.5 \omega_\text{H}$ for slightly higher spins than the first order solutions, and for determining how a higher order perturbation in spin translates to changes in our measure of error $\epsilon$.  It should be noted that the field line angular velocity of this solution is not uniform, as in the first order solution, but is a function of both $r$ and $\theta$.  

The fourth solution is a fully exact solution, described by Equation \ref{ExactSolnEqn}, with $h = \sin^3 \theta$ and $B_0 = 1$.  There are infinitely many possible choices for $h$.  Our selection is motivated by a desire for physical plausibility, which is most clearly seen by taking the Newtonian limit of the fields described.  In that limit, our choice of $h$ corresponds to magnetic and electric fields given by:
\begin{align}
\mathbf{B} &= -\frac{3 B_0}{r^2} \sin \theta \cos \theta \hat{r} - \frac{3 B_0}{a r} \cos \theta \hat{\phi}, \nonumber \\
\mathbf{E} &= \frac{3 B_0}{ar} \cos \theta \hat{\theta}.
\end{align} 
Here the vector components correspond to a standard orthonormal spherical basis.  We emphasize that the factor of $a$ above is from the definition of the fields; if the $a \rightarrow 0$ limit appropriate to a Newtonian transition were also taken in the electromagnetic fields they would diverge.  In the above form it is clear that our choice of $h = \sin^3 \theta$ was made to correspond to physically plausible fields, in the limited sense that they don't diverge anywhere in our domain; less careful choices generally diverge on the azimuthal axis.  This solution is most useful for estimating the potential precision of a given numerical grid for arbitrary black hole spin; we will refer to it as the \textit{HS3} solution when we compare it to our numerical results.

%%%%%%%%%%%%%%%%%%%%%%%%%%%%%%%%%%%%%%%%%%%%%%%%%%%%%%%%%%%%%%%%%%%%%%%%%%%%%%%%%%%%%%%%%%%%%%%%%%%%%%%%%%%%%%%%%%%%%%%%%%%%%%%%%%%%%%
\subsection{Computational Specifics}
%%%%%%%%%%%%%%%%%%%%%%%%%%%%%%%%%%%%%%%%%%%%%%%%%%%%%%%%%%%%%%%%%%%%%%%%%%%%%%%%%%%%%%%%%%%%%%%%%%%%%%%%%%%%%%%%%%%%%%%%%%%%%%%%%%%%%%

In this section we detail the computational specifics of our numerical methods, covering the grid resolutions used, domain boundaries and boundary conditions, initial conditions, and performance.  

The poloidal plane is divided into an $(r, \theta)$ grid for the vector potential $A_\phi$.  In the radial direction the grid extends from just inside the horizon to a radius $r_\text{max}$ outside the ergosphere.   The value of $r_\text{max}$ is dependent upon the location of the outer light surface, which is in turn dependent upon the choice of field line angular velocity $\Omega_\text{F}$ and black hole spin.  For small values of $\Omega_\text{F}$ (near 0) or black hole spin the outer light surface can be a very large distance from the black hole, so we artificially place a cap of $20 m$ on the maximum radius.  For large values of $\Omega_\text{F}$ (near $\omega_\text{H}$) and black hole spin the outer light surface approaches the ergosphere and $r_\text{max} \approx 2 m$ is used.  

It is critical to resolve the inner light surface, so for large values of $\Omega_\text{F}$ where the inner light surface approaches the horizon we use very small grid spacings in the radial direction near the horizon, then gradually relax that spacing as $r$ increases.  For small values of $\Omega_\text{F}$ the horizon and inner light surface are well separated and we use larger radial grid spacings near the horizon, then increase that spacing as $r$ extends past the ergosphere to relatively larger $r_\text{max}$ values.  In general around 400 grid squares of varied spacing in the radial direction are used to resolve the inner light surface and ergoregion and an additional $450$ of varied spacing are used to extend to a given $r_\text{max}$.  This yields a rough average of 850 grid squares in $r$ per full magnetosphere; exactly how many are used varies from magnetosphere to magnetosphere.  

In order to treat the value of $A_\phi$ on the horizon as an evolving entity, we extend three grid squares past the horizon towards the black hole.  We cannot extend any further as our implementation becomes anti-diffusive inside the horizon and more grid squares allow numerical instabilites to develop.  We are using Boyer-Lindquist coordinates, which are singular on the horizon.  This is not a difficulty, as we scale the magnetofrictional coefficient $\nu$ in Equation \ref{Eqn:NuEqn} by the magnitude of the poloidal magnetic field (see Appendix \ref{App:MFExpand}).  The coordinate singularity in that scaling cancels with the coordinate singularity in the transfield equation such that the horizon is well behaved.  As a solution is found the horizon naturally settles into a configuration consistent with the Znajek horizon condition (the force-free limit of the fast magnetosonic regularity condition, Equation \ref{ZnajekEqn}), justifying the usage of Boyer-Lindquist coordinates.

In the $\theta$ direction we extend from $\theta = 0$ on the azimuthal axis to $\theta = \pi/2$ on the equatorial plane using $200$ linearly spaced grid squares.  The equations in use do diverge on the azimuthal axis, a consequence of axisymmetry requiring that the magnetic field be purely radial there.  By using the azimuthal axis as a fixed boundary condition we enforce axisymmetry and avoid that divergence, and find that double precision arithmetic is sufficient for the grid squares immediately adjacent to the axis.  The other fixed boundary condition that we use is the equatorial plane, in keeping with the assumption of perturbing around a monopolar geometry (Section \ref{Sec:ModelAss}).  On the azimuthal axis we fix $A_\phi = 4$ and on the equatorial plane we fix $A_\phi = 3$, for a $\Delta A_\phi$ between them consistent with a monopole of unit weight ($A_\phi = \cos \theta$ in Equation \ref{SchwMonEqn}).  In principle any arbitrary range between the axis and equatorial plane could be used; we made our selection for convenience.         
     
We do not use fixed boundary conditions for the vector potential $A_\phi$ at either $r_\text{min}$ or $r_\text{max}$.  Instead, after every time step we shoot outwards to find a boundary that keeps $A_\phi$ smooth and use that boundary for the next iteration.  The reason for this is practical; although the axis and equatorial plane are fixed by symmetry, there is no restriction on $A_\phi$ for a given $r_\text{max}$.  The inner boundary could be fixed by applying the horizon regularity condition, but we chose to have the horizon evolve as a simple check that the numerical algorithm is converging properly.  

In order to calculate the derivatives in $A_\phi$ required to calculate the coordinate velocity $v^A$ of the fictitious plasma (Equations \ref{Eqn:Vexplicit1} and \ref{Eqn:Vexplicit2}) we use centered finite difference approximations appropriate to the local grid spacing.  One-sided finite difference derivatives in $A_\phi$ appropriate to an upwind differencing algorithm are then used to evolve the magnetofrictional advection equation (Equation \ref{Eqn:AdvecEqn}).  

Our initial conditions for each run were fairly simple.  Any smooth $A_\phi$ that decreased monotonically from the azimuthal axis to the equatorial plane could be used; we found no dependence upon initial conditions.  Lack of monotonicity caused ``spikes'' to develop in $A_\phi$ that required magnetic reconnection via numerical diffusion to dissipate, greatly extending computation time when the code remained stable.  To speed convergence we found it desirable to begin with the vector potential of the closest already calculated magnetosphere, ``close'' being defined by either slightly different black hole spin or field line angular velocity.  We do not directly use the toroidal field $\sqrt{-g} F^{\theta r}$, instead we propose a function of $A_\phi$ corresponding to the derivative with respect to $A_\phi$ of the square of the toroidal field (the left hand side of Equation \ref{Eqn:KinkFix}).  From the transfield equation it can be shown that this function must vanish on the axis and equatorial plane (at $A_\phi = 4$ or $A_\phi = 3$ using the boundary conditions described above) but is otherwise largely unrestricted.  Beyond increasing computation time we found no dependence upon initial choice of this function.  In order to decrease computation time we often found it desirable to use three neighboring magnetospheres with the same spin but different field line angular velocities to ``shoot'' an initial guess for this function for a given value of $A_\phi$.        

The overall number of time steps required to find a solution is sensitive to the accuracy of the initial guesses for both the vector potential and toroidal field, with the majority of computation time typically being spent reducing the kink at the inner light surface by finding a compatible toroidal field.  In general with good initial guesses for the fields and well-tuned parameters a desktop computer can find a magnetosphere at the $\epsilon = 1\%$ level (Section \ref{Sec:MeasuringFF}) in a matter of hours; a $ \epsilon \sim10\%$ level can often be achieved in less than an hour.  We have found that it is possible to significantly reduce computation time by bracketing the functional form of the derivative of the toroidal field  using a kind of two dimensional root-finding algorithm with initial guesses taken from adjacent magnetospheres, but that technique requires careful setup and tuning to avoid instabilities.  The na\"{i}ve method of kink reduction described in Section \ref{Sec:Kinks} is more robust, easier to implement, and while significantly slower is not prohibitively so for most purposes.      

%%%%%%%%%%%%%%%%%%%%%%%%%%%%%%%%%%%%%%%%%%%%%%%%%%%%%%%%%%%%%%%%%%%%%%%%%%%%%%%%%%%%%%%%%%%%%%%%%%%%%%%%%%%%%%%%%%%%%%%%%%%%%%%%%%%%%%
%%%%%%%%%%%%%%%%%%%%%%%%%%%%%%%%%%%%%%%%%%%%%%%%%%%%%%%%%%%%%%%%%%%%%%%%%%%%%%%%%%%%%%%%%%%%%%%%%%%%%%%%%%%%%%%%%%%%%%%%%%%%%%%%%%%%%%
%%%%%%%%%%%%%%%%%%%%%%%%%%%%%%%%%%%%%%%%%%%%%%%%%%%%%%%%%%%%%%%%%%%%%%%%%%%%%%%%%%%%%%%%%%%%%%%%%%%%%%%%%%%%%%%%%%%%%%%%%%%%%%%%%%%%%%

\section{Results} \label{Sec:Results}

We divide our results into four sections.  First we explore the general structure of the magnetospheres as a function of black hole spin parameter and field line angular velocity.  We then examine the rates of energy and angular momentum extraction.  Lastly we explore the numerical error of our solutions. 

%%%%%%%%%%%%%%%%%%%%%%%%%%%%%%%%%%%%%%%%%%%%%%%%%%%%%%%%%%%%%%%%%%%%%%%%%%%%%%%%%%%%%%%%%%%%%%%%%%%%%%%%%%%%%%%%%%%%%%%%%%%%%%%%%%%%%%
\subsection{Field Line Structure}
%%%%%%%%%%%%%%%%%%%%%%%%%%%%%%%%%%%%%%%%%%%%%%%%%%%%%%%%%%%%%%%%%%%%%%%%%%%%%%%%%%%%%%%%%%%%%%%%%%%%%%%%%%%%%%%%%%%%%%%%%%%%%%%%%%%%%%

\begin{figure*}[p]
     \includegraphics[height=0.85\textheight,keepaspectratio]{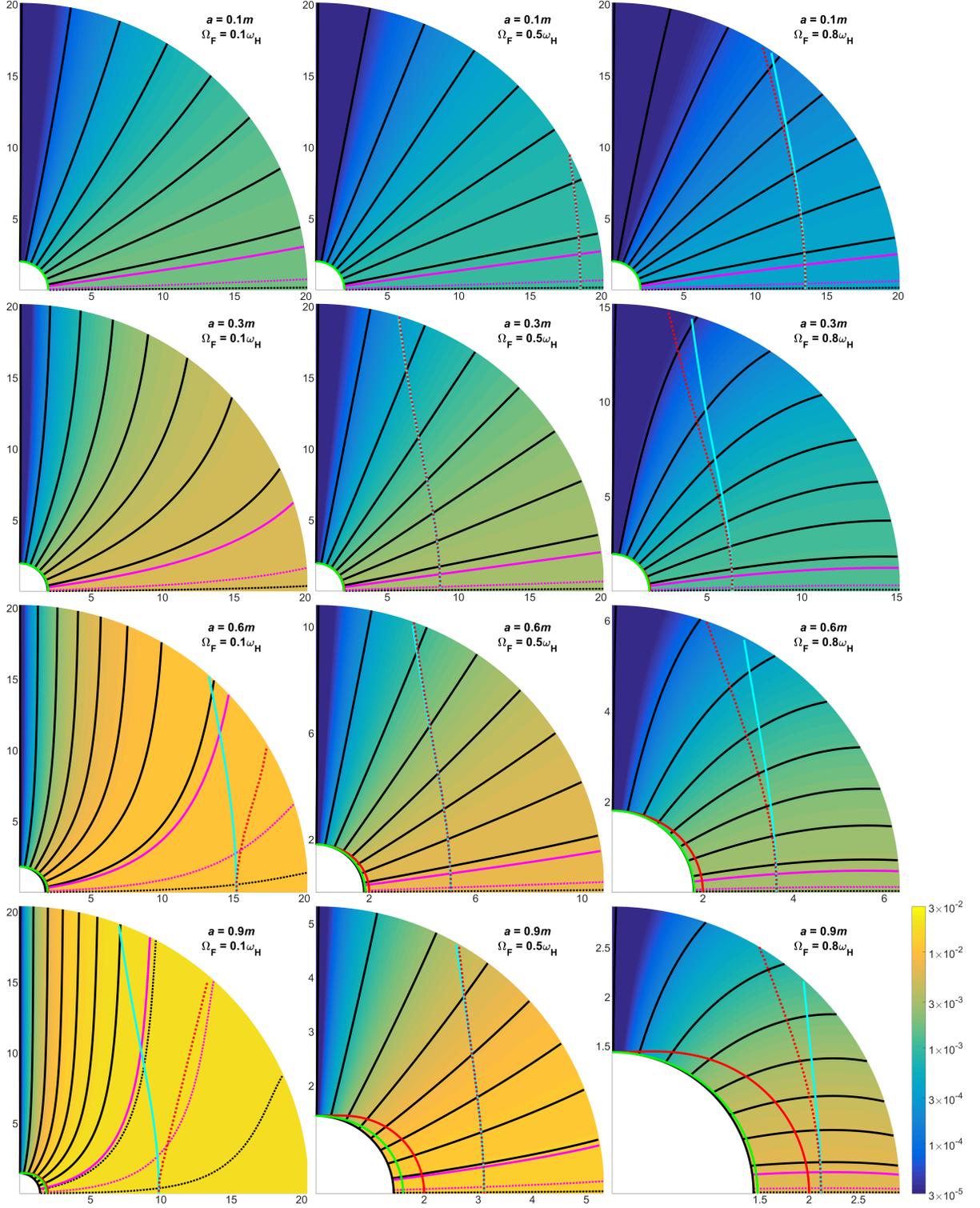}
     \caption{The structure of poloidal magnetic field lines for various values of black hole spin and field line angular velocity.  The magnitude of the outward conserved angular momentum flux per unit field line ($L$ in Equation \ref{EandLEquation}) is colored, scaled to $\Delta A_\phi = 1$ between the axis and the equatorial plane ($B_0 = 1$ in Equation \ref{SchwMonEqn}).   The inner light surface (green), ergosphere (red), horizon (black), monopole separation surface (cyan), and calculated separation surface (dotted red) are overplotted.  80\% of the total extracted energy flows outward between the azimuthal axis and the magnetic field line drawn in solid magenta; 95\% flows between the dotted magenta field line and the axis.  The eight black field lines are spaced according to the magnitude of the radial magnetic field on the horizon; if they fall outside the 80\% line they are dotted.}
			\label{fig:FieldLines2}
\end{figure*}

\begin{figure*}[p]
     \includegraphics[height=0.85\textheight,keepaspectratio]{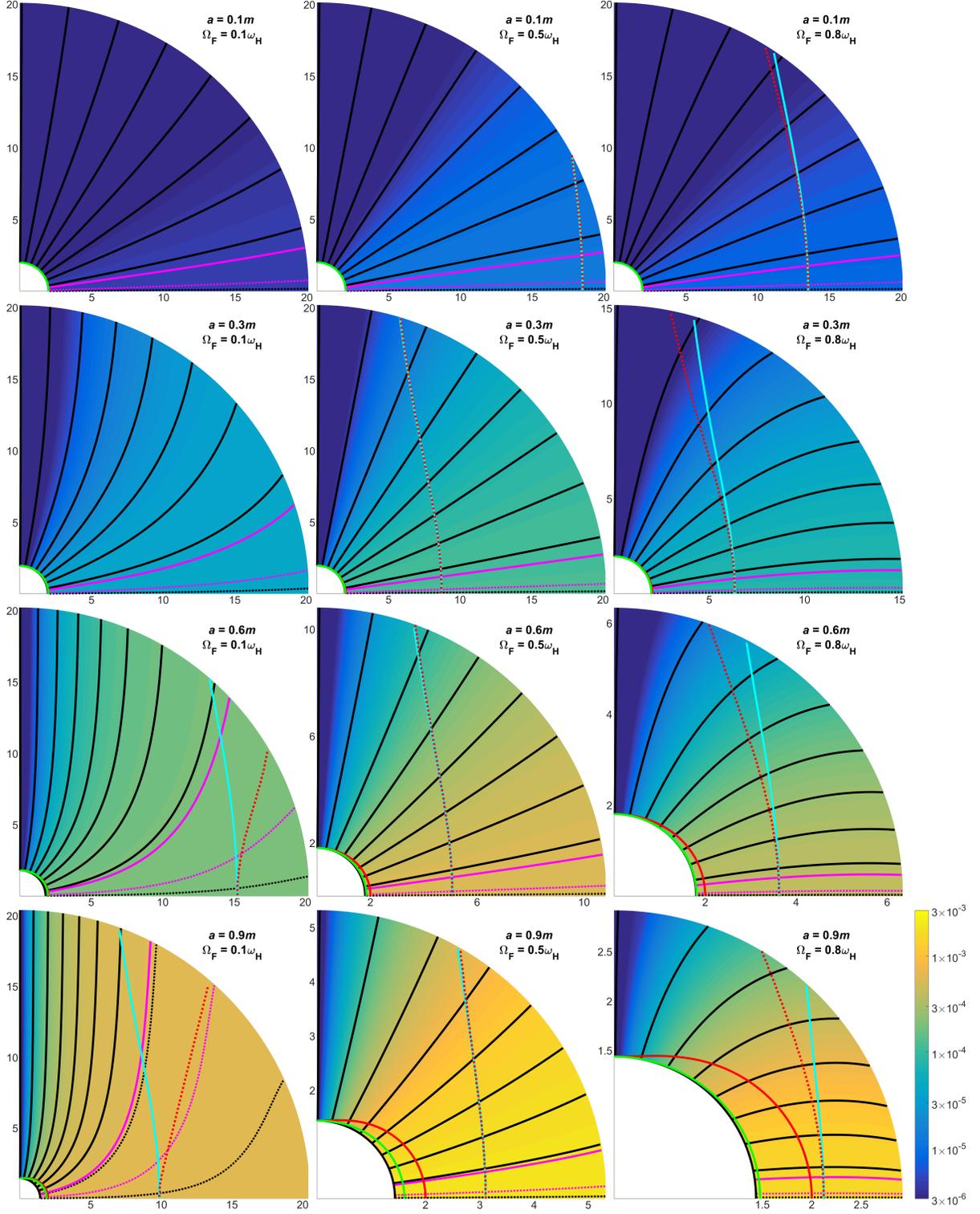}
     \caption{The same structure of poloidal magnetic field lines as in Figure \ref{fig:FieldLines2}, but with the magnitude of the outward conserved energy flux per unit field line ($E$ in Equation \ref{EandLEquation}) colored.  The inner light surface (green), ergosphere (red), horizon (black), monopole separation surface (cyan), and calculated separation surface (dotted red) are overplotted.  80\% of the total extracted energy flows outward between the azimuthal axis and the magnetic field line drawn in solid magenta; 95\% flows between the dotted magenta field line and the axis.  The eight black field lines are spaced evenly on the horizon; if they fall outside the 80\% line they are dotted.}
			\label{fig:FieldLines1}
\end{figure*}

In this section we explore the behavior of poloidal magnetic field lines. We are limited to regions near to the black hole by our model assumptions and numerical techniques (Sections \ref{Sec:ModelAss} and \ref{Sec:Kinks}).  However even over that limited domain interesting behaviors emerged, as shown in Figures \ref{fig:FieldLines2} and \ref{fig:FieldLines1}.  

Figure \ref{fig:FieldLines2} plots poloidal magnetic field lines using spacing on the horizon corresponding to the strength of the radial magnetic field there (using $|B^r| \sim \csc \theta A_{\phi, \theta}$ appropriate to ingoing coordinates); denser field lines imply greater magnetic field strength.  The first and last field line do not lie on the axis or equatorial plane, as those are fixed, but instead lie one grid square inwards on the horizon.  The shading measures conserved momentum flux per unit field line ($L$ in Equation \ref{EandLEquation}), which is equivalent to the toroidal field $\sqrt{-g} F^{\theta r}$ multiplied by $4\pi$.  

Figure \ref{fig:FieldLines1} plots poloidal magnetic field lines using linear spacing in $\theta$ on the horizon.  The first and last field line again lie one grid square inwards on the horizon.  The shading measures conserved energy flux per unit field line ($E$ in Equation \ref{EandLEquation}).

The first thing to note is that as expected for very small black hole spin ($a \approx 0.1m$) all values of field line angular velocity yield magnetospheres that are almost indistinguishable from a Schwarzschild monopole (Equation \ref{SchwMonEqn}, noting that ``monopole'' refers only to the poloidal plane).  For slightly larger values of spin ($a \approx 0.3m$), however, noticeable deviations from a monopolar structure rapidly emerge for both low and high field line angular velocities.  As spin increases beyond $a \approx 0.3 m$, only the $\Omega_\text{F} \approx 0.5 \omega_\text{H}$ solutions remain roughly monopolar. 

Low field line angular velocity solutions bend towards the azimuthal axis, with the strength of that bending increasing as black hole spin increases.  For large spin ($a \approx 0.9m$) this bending can be severe, with the majority of poloidal magnetic field lines becoming nearly parallel to the azimuthal axis at a distance $r = 20m$.  The source of this bending is most easily seen in Figure \ref{fig:FieldLines2}; on the horizon the radial magnetic field is strong on the azimuthal axis and weak near the equatorial plane, while the toroidal magnetic field is strong on the equatorial plane and weak on the azimuthal axis.  This discrepancy acts to naturally wind up the magnetic field and the magnetosphere self-collimates into a jet-like structure.  The discrepancy becomes very large at high spin; for $a=m$ (not shown) the horizon radial magnetic field strength on the axis can be $\sim10$ times that on the equator.  For larger values of $\Omega_\text{F}$ (implying larger electric fields as viewed by distant observers, but not necessarily others such as ZAMOs) the toroidal field becomes much smaller, radial magnetic field strength becomes almost uniform on the horizon, and the field lines begin to bend toward the equatorial plane.

There are two separation surfaces shown in the figures (the separation surface may be defined as the point where $\alpha' = 0$, with the prime denoting differentiation along magnetic field lines); one for a monopolar geometry where $A_\phi$ is purely a function of $\theta$ and one corresponding to the numerically calculated $A_\phi$.  For low field line angular velocities and high spins the separation surface moves away from the horizon at higher latitudes, while for high field line angular velocities it moves towards the horizon. 

Figure \ref{fig:FieldLines1} makes it clear that significantly more energy is extracted along field lines near the equator for all values of field line angular velocity, but low $\Omega_\text{F}$ and high spin magnetospheres redirect most of that extracted energy towards the azimuthal axis.  We measure this behavior in the next section. 

%%%%%%%%%%%%%%%%%%%%%%%%%%%%%%%%%%%%%%%%%%%%%%%%%%%%%%%%%%%%%%%%%%%%%%%%%%%%%%%%%%%%%%%%%%%%%%%%%%%%%%%%%%%%%%%%%%%%%%%%%%%%%%%%%%%%%%
\subsection{Energy Extraction} \label{Sec:EExtraction}
%%%%%%%%%%%%%%%%%%%%%%%%%%%%%%%%%%%%%%%%%%%%%%%%%%%%%%%%%%%%%%%%%%%%%%%%%%%%%%%%%%%%%%%%%%%%%%%%%%%%%%%%%%%%%%%%%%%%%%%%%%%%%%%%%%%%%%

In this section we explore the rate of black hole energy extraction.  Energy flux is conserved along magnetic field lines, so the rate of black hole energy extraction is most easily calculated on the horizon.  If we define $P$ as the power (energy per unit time as measured by a distant observer) leaving the horizon, we find (cf. \citet{LGATN2014}):
\begin{align} \label{Eqn:PRaw}
P &= \int_{r+} T^r{}_t \sqrt{-g} d \theta d \phi \nonumber \\
&= \frac{1}{2}\int_{r+} \Omega_\text{F} A_{\phi, \theta} \sqrt{-g} F^{\theta r} d\theta. 
\end{align}   
As this is evaluated on the horizon, we may use the Znajek horizon condition (Equation \ref{ZnajekEqn}) to find:
\begin{equation} \label{PEqn}
P = \frac{1}{2} Q \left(1 - Q \right) \frac{a^2}{r_+^2 + a^2} \int_0^{\pi} \frac{1}{\Sigma} A_{\phi, \theta}^2 \sin \theta d \theta . 
\end{equation}
Here $Q$ is a unitless scaling of the field line angular velocity to the horizon's angular velocity; $\Omega_\text{F} = Q \omega_\text{H}$ (for energy extraction to take place we must have $0 < Q < 1$).  In CGS units, the power $P$ becomes:
\begin{equation} \label{P2Eqn}
P = 5.2 \times 10^{19} \cdot \chi \cdot r_{x*}^4 \frac{B_x^2}{\text{G}^2} \frac{m^2}{M_\odot^2}  \frac{\text{erg}}{\text{s}},
\end{equation} 
where
\begin{equation} \label{ChiEqn}
\chi =  \frac{1}{2} Q \left(1 - Q \right) \frac{a_*^2}{\left(r_{+*}^2 + a_*^2 \right)} \int_0^\pi  \frac{A_{\phi, \theta}^2 \sin \theta}{r_{+*}^2 + a_*^2 \cos^2 \theta} d \theta. 
\end{equation} 
Here $a_*$ and $r_{+*}$ are dimensionless measures of black hole spin and horizon radius; $a = a_* m$ and $r_+ = r_{+*} m$.  The quantity $\chi$ is a dimensionless measure of the rate of black hole energy extraction that varies from magnetosphere to magnetosphere.  The quantities $B_x$ and $r_{x*}$ are measures of monopolar magnetic field strength, in a sense that we now explore by considering the Newtonian limit.  

In the Newtonian limit, a vector potential $A_\phi = B_0 \cos \theta$ corresponds to a monopolar magnetic field that is given by:
\begin{equation}  \label{flatmonopoleeqn}
\mathbf{B} = \frac{B_0}{r^2} \hat{r}.
\end{equation} 
Here the vectors correspond to a standard orthonormal spherical basis.  The weighting of $B_0$ on $A_\phi$ may therefore be interpreted as a measure of the monopolar magnetic field strength at a specified radius.  For convenience our numerical solutions for $A_\phi$ assume unit spacing between the azimuthal axis and the plane ($B_0 = 1$), but any arbitrary weighting would yield identical results.  The quantities $B_x$ and $r_{x*}$ are measures of that arbitrary weighting; $B_x$ is the field strength in Gauss of a Newtonian monopole at dimensionless radius $r_{x*}$.  For example, $B_x = 100 \text{G}$ and $r_{x*} = 20$ would correspond to a monopolar magnetic field strength of 100 Gauss at a distance $r = 20m$ from a black hole of mass $m$.  Some caution should be taken in applying this interpretation, however.  None of our numerical solutions are exactly monopolar, and the strength of the magnetic field is an observer dependent quantity that does not have a simple translation from a flat space value to the spacetime of a rotating black hole.  The above interpretation of $B_x$ and $r_{x*}$ is made purely for simplicity and convenience, and in general should be taken to be nothing more than a rough estimate.

\begin{figure}[ht]
     \includegraphics[width=\columnwidth,clip=true]{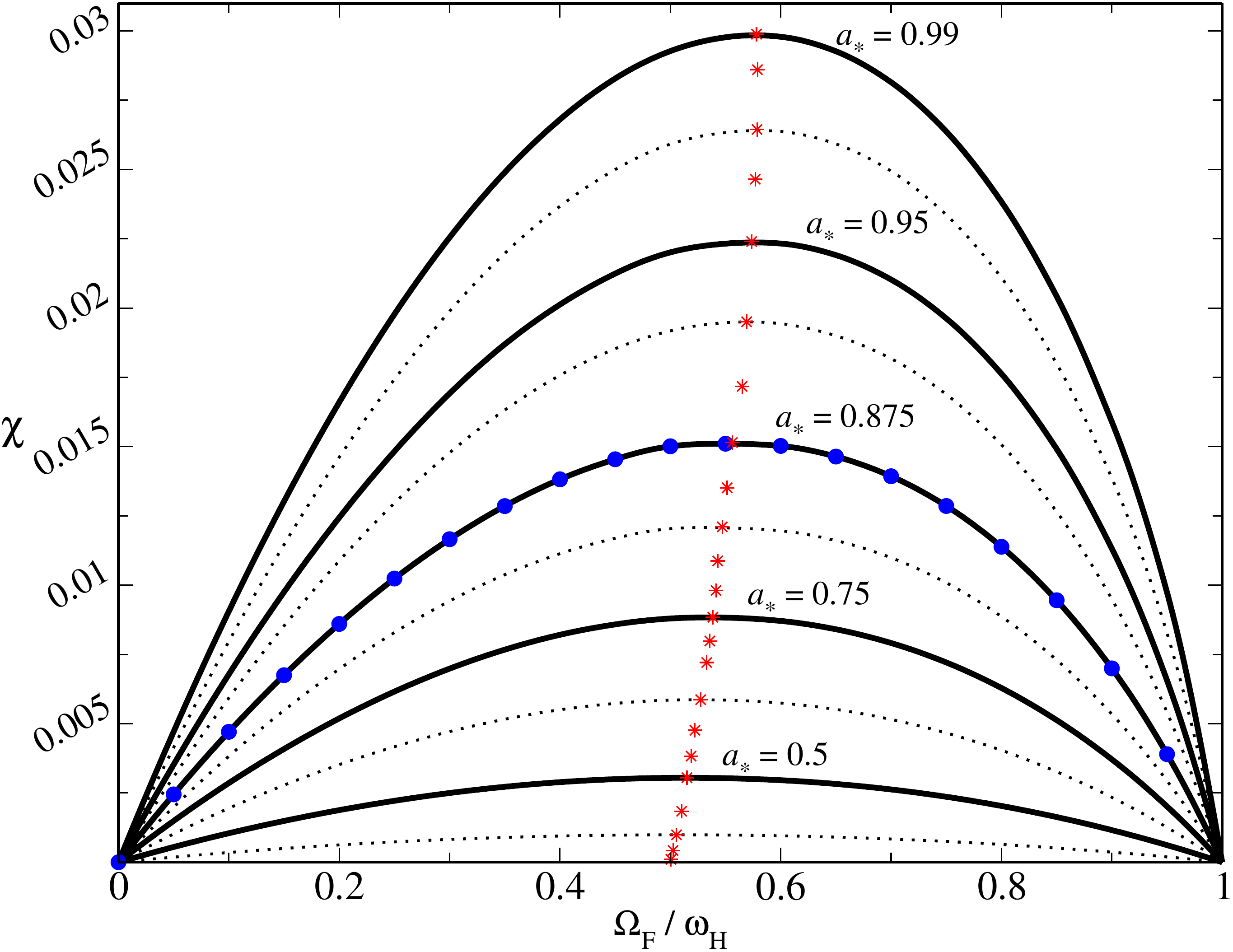}
     \caption{The rate of black hole energy extraction for select magnetospheres; $\chi$ is defined in Equation \ref{ChiEqn}.  For each value of black hole spin 20 magnetospheres were calculated, indicated for $a_*=0.875$ with blue dots.  The approximate maximum values of $\chi$ for various spin values are indicated with red stars; they corresponding to the 23 spin parameters $a_*$ 0.1, 0.2, 0.3, 0.4, 0.5, 0.55, 0.6, 0.65, 0.7, 0.725, 0.75, 0.775, 0.8, 0.825, 0.875, 0.9, 0.925, 0.95, 0.965, 0.975, 0.985, and 0.99.  For small spin the maximum value of $\chi$ occurs near $\Omega_\text{F} = 0.5 \omega_\text{H}$; for $a_* = 0.99$ the maximum occurs near $\Omega_\text{F} = 0.58 \omega_\text{H}$.  For extremal spin ($a_* = 1$, not shown) the rate of energy extraction peaks slightly above $\chi = 0.033$.}  
	   \label{fig:Energy}
\end{figure}

In Figure \ref{fig:Energy} we plot the value of $\chi$ as a measure of the rate of black hole energy extraction for select values of black hole spin and field line angular velocity.  We suppress spins greater than $a=0.99m$ as showing them would compress the curves of smaller spin that are of greater interest.  

For a given value of $\Omega_\text{F}$, we note that increasing black hole spin always increases the rate of energy extraction but that changes in field line angular velocity can have a much larger effect than changing spin.  The peak energy extracted for a given spin occurs at $\Omega_\text{F} \approx 0.5 \omega_\text{H}$ for low spin and approaches $0.6 \omega_\text{H}$ at high spin.  The maximum energy extracted spans two orders of magnitude; $\chi_\text{Max} = 0.03$ for $a \approx 0.99 m$, $\chi_\text{Max} = 0.003$ for $ a \approx 0.5 m$, and $\chi_\text{Max} = 0.0003$ for $a \approx 0.2m$.

To a very good approximation, the functional form of $\chi(a)$ for fixed field line angular velocities $\Omega_\text{F} = 0.1 \omega_\text{H}$, $\Omega_\text{F} = 0.95 \omega_\text{H}$, and value for maximum energy extraction $\Omega_\text{F Max}$ are given by:
\begin{align} \label{fitChiEqn}
\chi_{0.1 \omega_\text{H}} &\approx 3.3 \times 10^{-3} \left(1.8 a_*^{2.5} + 1.1 a_*^{17}\right), \nonumber \\
\chi_{0.95 \omega_\text{H}} &\approx 3.3 \times 10^{-3} \left(1.5 a_*^{3.1} + 1.8 a_*^{17}\right), \nonumber \\
\chi_{\text{Max}} &\approx 1.0\times10^{-2} \left(1.9 a_*^{2.6} + 1.3 a_*^{14}\right).
\end{align}
In this form it is clear that for a given spin the maximum rate of energy extraction corresponding to $\Omega_\text{F Max}$ is roughly three times that of the lower values $\Omega_\text{F} \approx 0.1 \omega_\text{H}$ or $\Omega_\text{F} \approx 0.95 \omega_\text{H}$, an effect that can be much larger than changes in spin.  

\begin{figure}[ht]
     \includegraphics[width=0.47\textwidth]{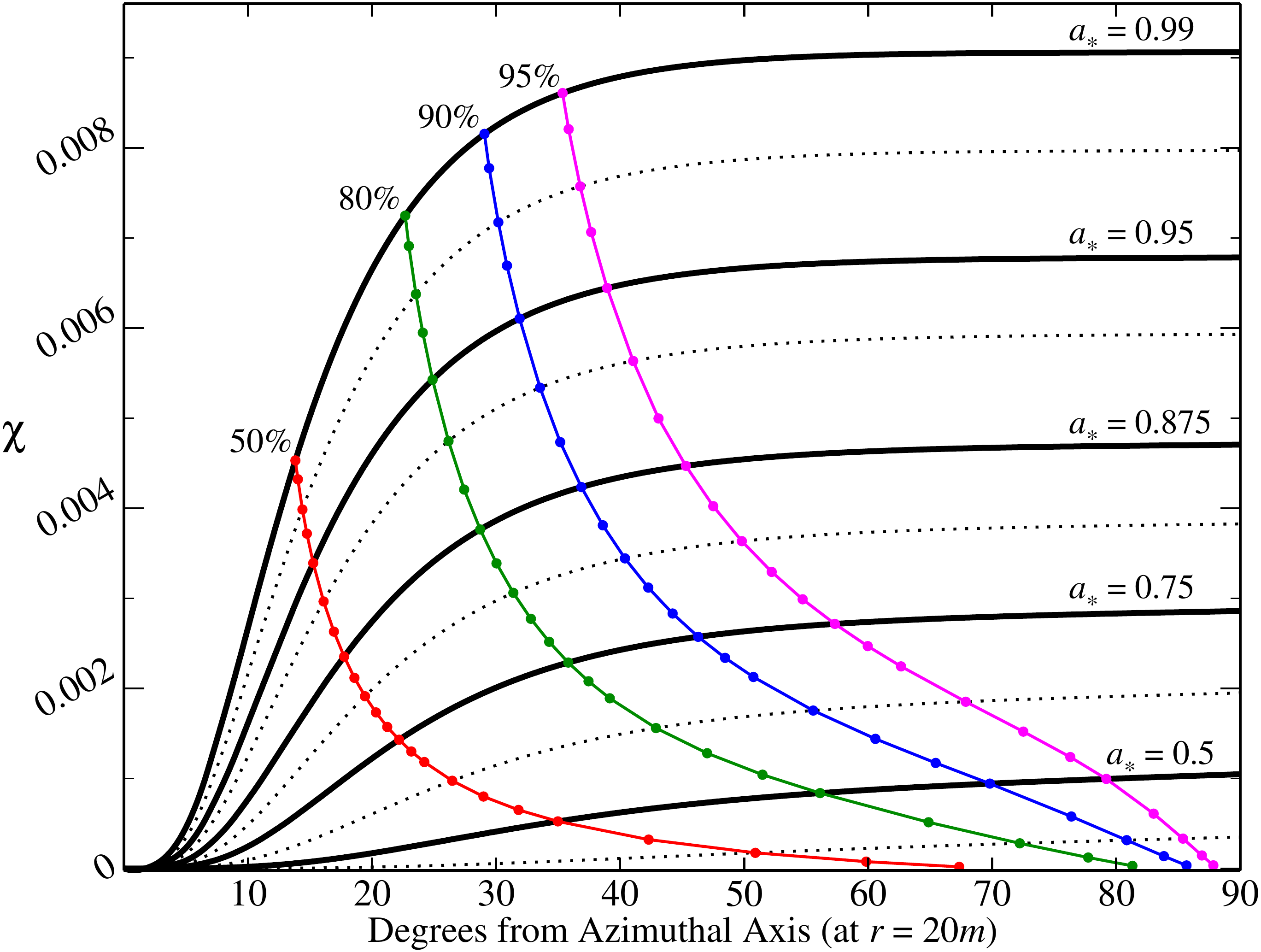}
     \caption{The rate of black hole energy extracted as a function of the angle of a spherical cone at a radius of $20m$ (in both upper and lower hemispheres) for a field line angular velocity $\Omega_\text{F} = 0.1 \omega_\text{H}$.  The angles through which 50\%, 80\%, 90\%, and 95\% of the total extracted energy escapes are overplotted.}  
	   \label{fig:EnergyAngle}
\end{figure}

For low field line angular velocities the magnetic field bends towards the azimuthal axis, as shown in Figures \ref{fig:FieldLines2} and \ref{fig:FieldLines1}.  So while most energy is extracted on the horizon near the equatorial plane, a short distance from the horizon a large fraction of it ends up flowing outward along the azimuthal axis.  In order to explore this behavior, we calculated the energy escaping through a spherical cone (both upper and lower hemispheres) at a radius $r = 20m$ for $\Omega_\text{F} = 0.1 \omega_\text{H}$.  The results are plotted in Figure \ref{fig:EnergyAngle}.  For high spins ($a_* \approx 0.85$ and greater), over 95\% of the extracted energy escapes through a $45^\circ$ cone, and 80\% escapes through a cone of less than $30^\circ$.  This indicates that for high spin and low field line angular velocities a black hole magnetosphere is naturally inclined to extract energy via jet-like structures aligned with the rotational axis of the black hole.

%%%%%%%%%%%%%%%%%%%%%%%%%%%%%%%%%%%%%%%%%%%%%%%%%%%%%%%%%%%%%%%%%%%%%%%%%%%%%%%%%%%%%%%%%%%%%%%%%%%%%%%%%%%%%%%%%%%%%%%%%%%%%%%%%%%%%%
\subsection{Angular Momentum Extraction}
%%%%%%%%%%%%%%%%%%%%%%%%%%%%%%%%%%%%%%%%%%%%%%%%%%%%%%%%%%%%%%%%%%%%%%%%%%%%%%%%%%%%%%%%%%%%%%%%%%%%%%%%%%%%%%%%%%%%%%%%%%%%%%%%%%%%%%

\begin{figure}[ht]
     \includegraphics[width=0.47\textwidth]{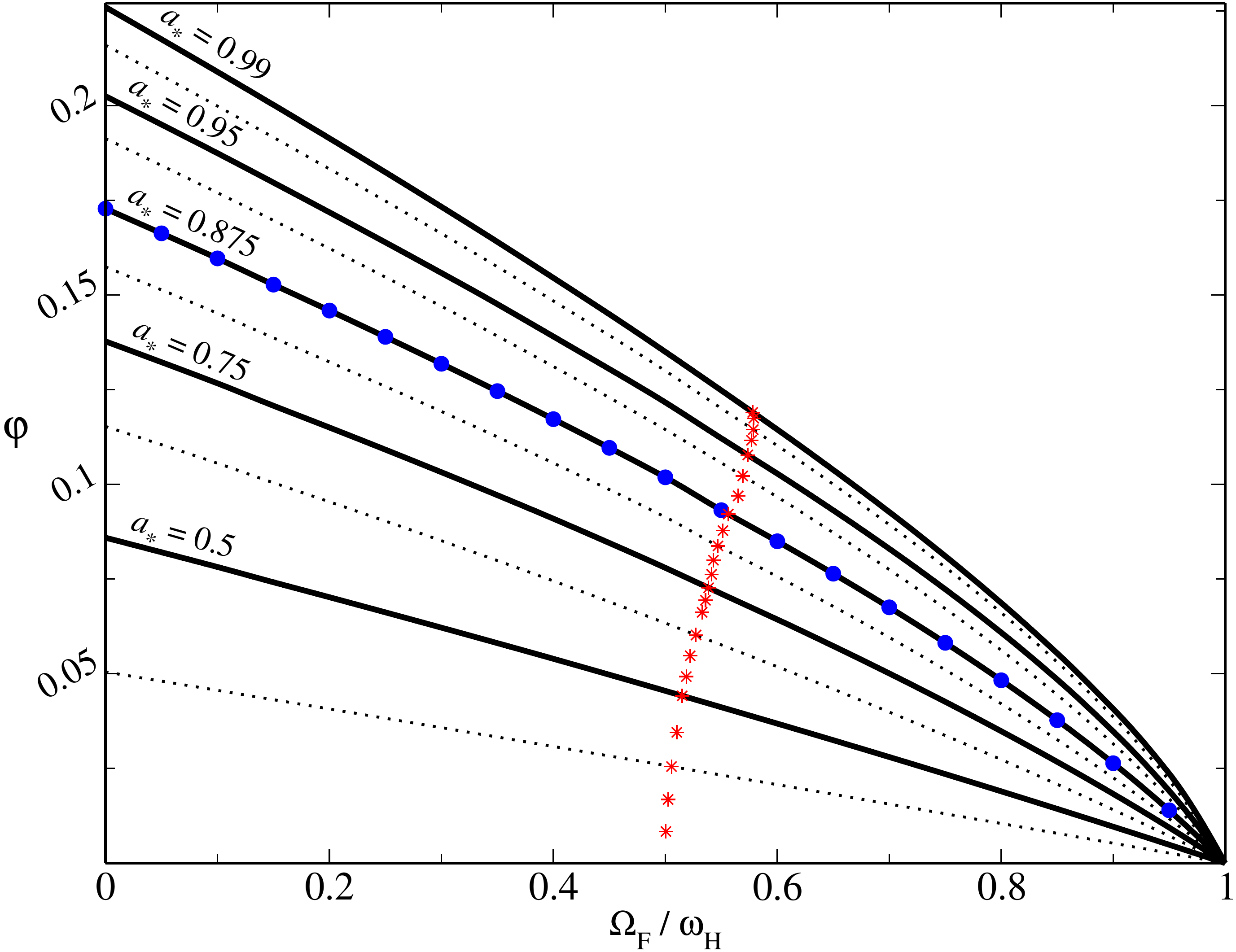}
     \caption{The rate of black hole angular momentum extraction for the selection of magnetospheres used in Figure \ref{fig:Energy}; $\varphi$ is defined in Equation \ref{VarphiEqn}.  For each spin the angular momentum extracted at the maximum rate of energy extraction is indicated by red stars.  The blue dots indicate the 20 values of field line angular velocity used for the $a_* = 0.875$ case.  The red stars indicate the 23 different values of spin listed in Figure \ref{fig:Energy}.}  
	   \label{fig:Momentum}
\end{figure}

In this section we explore the rate of black hole angular momentum extraction.  As with energy flux, angular momentum flux is conserved along magnetic field lines, and is most easily calculated on the horizon by exploiting the Znajek horizon condition.  If we define $K$ as the rate of angular momentum extraction, we find:
\begin{align}
K &= -\int_{r+} T^r{}_\phi \sqrt{-g} d \theta d \phi \nonumber \\
&= \frac{1}{2}\int_{r+} A_{\phi, \theta} \sqrt{-g} F^{\theta r} d\theta. 
\end{align}   
Note that this only differs from the rate of energy extraction (Equation \ref{Eqn:PRaw}) by a factor of field line angular velocity $\Omega_\text{F}$, consistent with the conserved energy $E$ and angular momentum $L$ per unit flux tube differing by the same factor (Equation \ref{EandLEquation}).  In CGS units $K$ reduces to: 
\begin{equation} 
K = 2.6 \times 10^{14} \cdot \varphi \cdot r_{x*}^4 \frac{B_x^2}{\text{G}^2} \frac{m^3}{M_\odot^3} \text{erg}, \label{Eqn:LwithBx}
\end{equation}
where
\begin{equation}
\varphi =  \frac{1}{2} \left(1 - Q \right) a_* \int_0^\pi  \frac{A_{\phi, \theta}^2 \sin \theta}{r_{+*}^2 + a_*^2 \cos^2 \theta} d \theta. \label{VarphiEqn}
\end{equation} 
The primary structural difference between these expressions and those for the rate of energy extraction (Equations \ref{P2Eqn} and \ref{ChiEqn}) is the fact that $\varphi$ does not vanish for $Q = 0$.  This is a statement that it is possible to spin down a black hole without extracting energy.  In Figure \ref{fig:Momentum} we plot the value of $\varphi$ as a measure of the rate of black hole angular momentum extraction for the same values of black hole spin and field line angular velocity that were used to plot the rate of energy extraction.

To a very good approximation, the functional form of $\varphi(a)$ for fixed field line angular velocities $\Omega_\text{F} = 0.1 \omega_\text{H}$, $\Omega_\text{F} = 0.95 \omega_\text{H}$, and value for maximum energy extraction $\Omega_\text{F Max}$ are given by:
\begin{align} \label{fitVarphiEqn}
\varphi_{0.1 \omega_\text{H}} &\approx 1.7\times10^{-2} \left(9.7 a_*^{1.1} + 2.8 a_*^{8.0} \right), \nonumber \\
\varphi_{0.95 \omega_\text{H}} &\approx 1.3\times10^{-3} \left(9.4 a_*^{1.3} + 9.7 a_*^{9.3} \right), \nonumber \\
\varphi_{\text{Max}} &\approx 1.0\times10^{-2} \left(9.3 a_*^{1.1} + 2.8 a_*^{6.6}\right).
\end{align}
In this form it is clear that the $\Omega_\text{F} = 0.1 \omega_\text{H}$ solutions extract nearly twice as much angular momentum as the solutions that maximize the rate of energy extraction.  It is also clear that the $\Omega_\text{F} = 0.95 \omega_\text{H}$ solutions extract around 10\% of the angular momentum of the $\Omega_\text{F} = 0.1 \omega_\text{H}$ solutions, while from Equation \ref{fitChiEqn} we note that these solutions extract roughly the same amount of energy.  This indicates that high $\Omega_\text{F}$ solutions can extract energy from a black hole for a much longer period of time than lower $\Omega_\text{F}$ solutions can, as it will take longer for high $\Omega_\text{F}$ magnetospheres to spin down the black hole.

%%%%%%%%%%%%%%%%%%%%%%%%%%%%%%%%%%%%%%%%%%%%%%%%%%%%%%%%%%%%%%%%%%%%%%%%%%%%%%%%%%%%%%%%%%%%%%%%%%%%%%%%%%%%%%%%%%%%%%%%%%%%%%%%%%%%%%
\subsection{Numerical Error Estimation} \label{Sec:NumErrRes}
%%%%%%%%%%%%%%%%%%%%%%%%%%%%%%%%%%%%%%%%%%%%%%%%%%%%%%%%%%%%%%%%%%%%%%%%%%%%%%%%%%%%%%%%%%%%%%%%%%%%%%%%%%%%%%%%%%%%%%%%%%%%%%%%%%%%%%

In this section we explore the numerical error of our solutions.  In Section \ref{Sec:AnalyticComp} we discussed four analytic solutions that can provide useful comparisons with our numerical results, most especially in determining the reliability of the $\epsilon = 1\%$ error level we have set (discussed in Section \ref{Sec:MeasuringFF} and Appendix \ref{PercentAppendix}).   

\begin{figure}[ht]
     \includegraphics[width=0.47\textwidth]{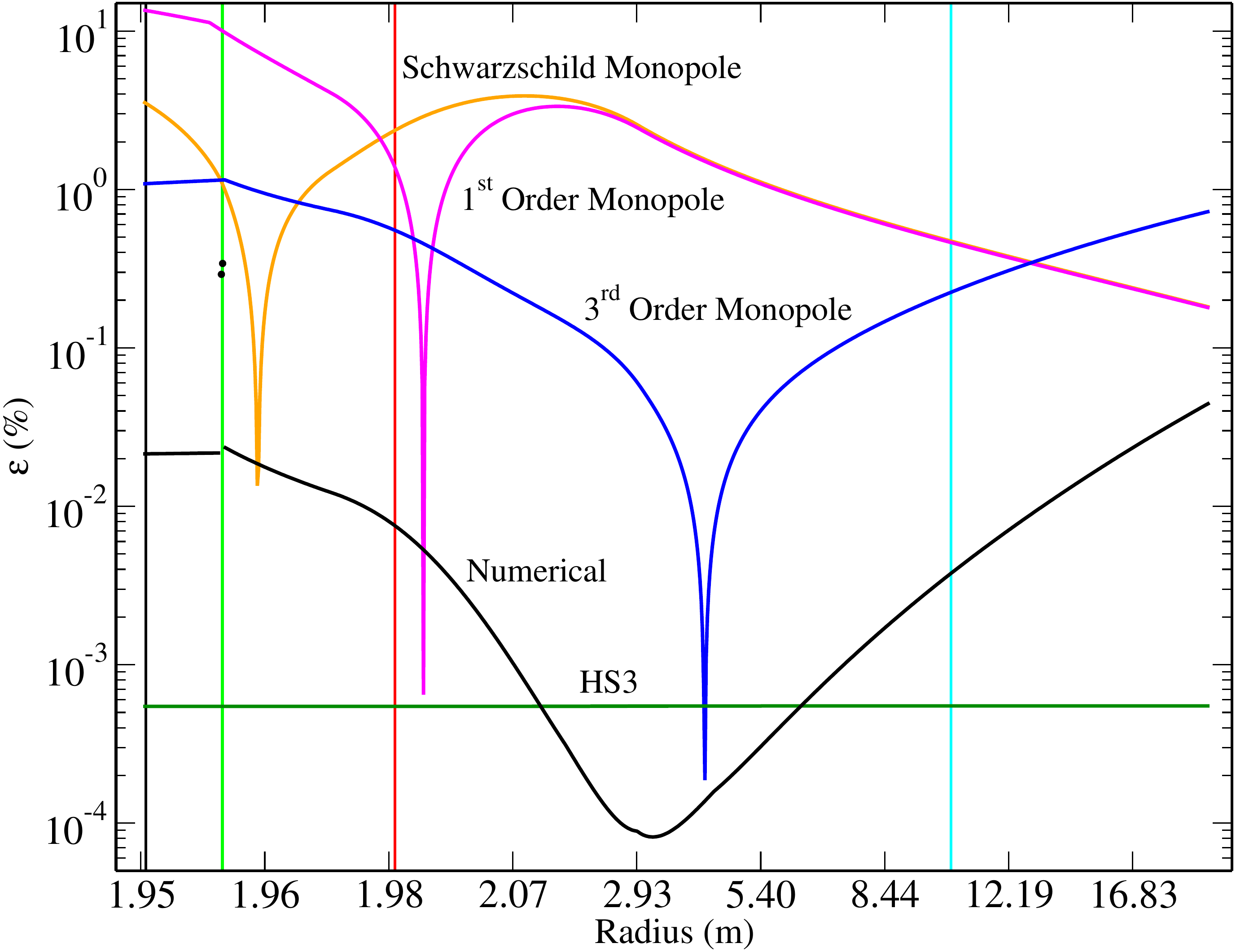}
     \caption{The percent error along a $\theta = 45^\circ$ slice for $a = 0.3m$ and $\Omega_\text{F} = 0.5 \omega_\text{H}$; the radial spacing corresponds to uniform spacing in the numerical grid.  The horizon (H), inner light surface (ILS), ergosphere (E), and separation point (SP) are indicated by vertical lines.  With the exception of the two grid squares immediately adjacent to the inner light surface the numerical solution is significantly better than the perturbed solutions.  The various solutions used are listed in Section \ref{Sec:AnalyticComp}}  	   \label{fig:ErrorConstTheta}
\end{figure}

The three perturbed monopole and exact \textit{HS3} solutions have varying regions of applicability.  We examine the numerical solution for $a=0.3m$ and $\Omega_\text{F} = 0.5 \omega_\text{H}$ so that we can reasonably compare all four solutions simultaneously.  For convenience we begin by comparing them along a slice of constant $\theta = 45^\circ$ from just inside the horizon to $r_\text{max} = 20m$ (Figure \ref{fig:ErrorConstTheta}).  Along that slice it is apparent that the Schwarzschild monopole and first order perturbed monopole are very similar, with both typically having errors of less than 5\%.  The third order perturbed solution does better in general, with an error of around 1\% near the inner light surface and inside the ergosphere.  Aside from the two grid squares immediately adjacent to the inner light surface the numerical solution has an error of less than 0.05\% across the entire slice, and is sometimes ``better'' than the \textit{HS3} solution's error of around 0.0005\%.  We place quotes around ``better'' because the error of the \textit{HS3} solution is a rough measure of the precision of our numerical grid and derivatives; significantly exceeding $\sim 0.0005\%$ likely involves unjustified precision.  Slices along different $\theta$ values as well as different spin parameters and field line angular velocities yield qualitatively similar results.  

\begin{figure}[ht]
     \includegraphics[width=0.47\textwidth]{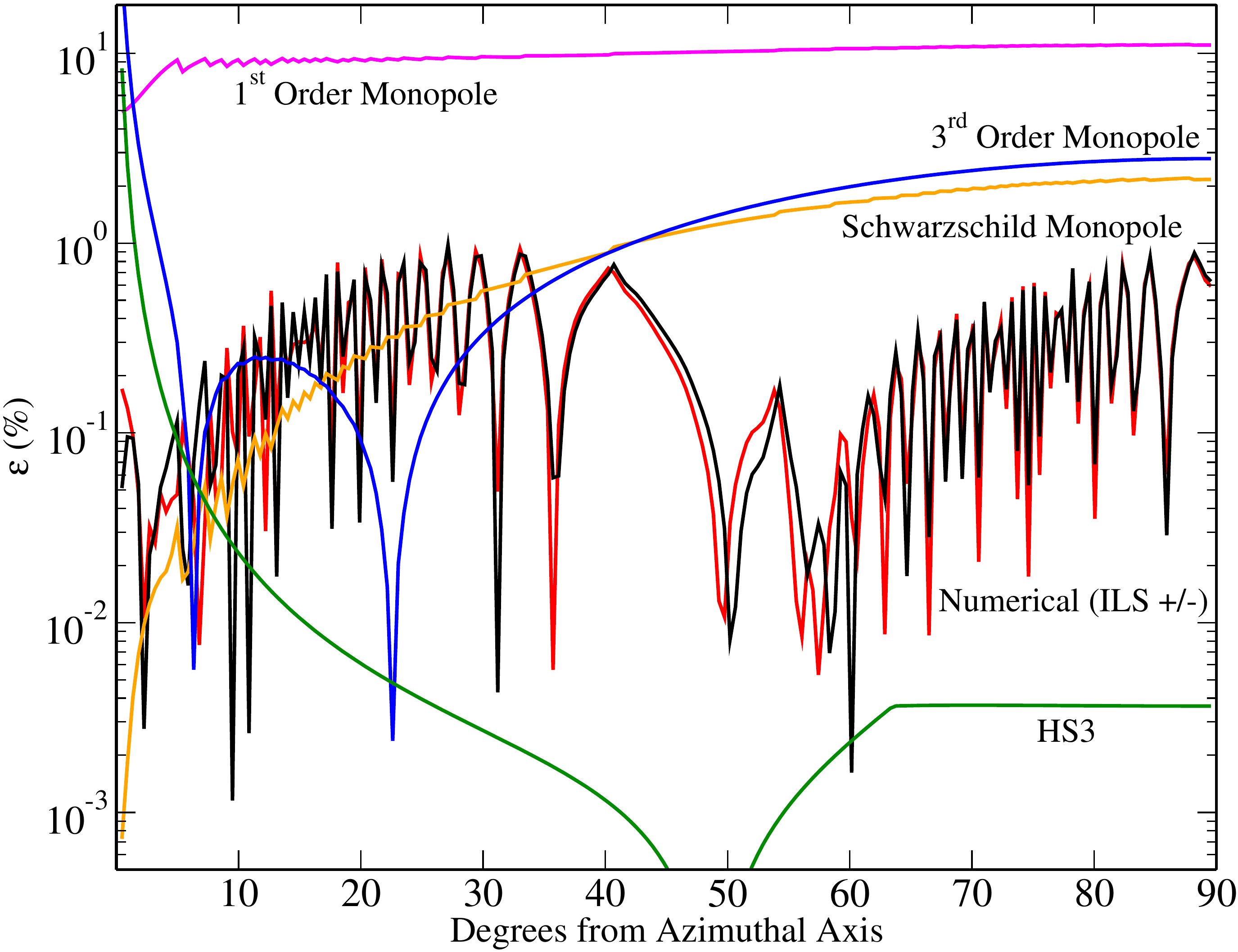}
     \caption{The percent error along the inner light surface for $a = 0.3m$ and $\Omega_\text{F} = 0.5 \omega_\text{H}$.  The grid squares on either side of the inner light surface (red and black) have almost identical error; reducing the spikes in that error to the 1\% threshold we have set is what takes the most computational time.  Despite those spikes the numerical solution still generally does as good or better than the perturbed solutions.  The upticks in the third order monopole and exact \textit{HS3} solutions near $\theta = 0$ are from the conversion of the toroidal field to a differentiated function of $A_\phi$.  Most of the terms in the transfield equation are very small there, so numerical errors in that conversion are amplified.}  
	   \label{fig:ErrorLS}
\end{figure}

In order to examine the increase in error along the inner light surface we plot the percent error of the numerical solution in the two grid squares immediately adjacent to the inner light surface in Figure \ref{fig:ErrorLS}.  This is done for all values of $\theta$, again for $a = 0.3m$ and $\Omega_\text{F} = 0.5 \omega_\text{H}$.  Even though substantial portions of the inner light surface are well below the 1\% level, there are spikes up to 1\%.  Reducing those spikes is what takes the largest amount of computational time.  The perturbed solutions are either comparable to or greatly exceed the error in the numerical solutions along the entirety of the inner light surface. The \textit{HS3} solution is typically 10-100 times better than any other solution.  The exception is near the azimuthal axis where the \textit{HS3} solution's field line angular velocity diverges.  Note that the error of the \textit{HS3} solution does not also diverge, as the combined equations are well behaved there; the error increase comes from amplification of the error in taking numerical derivatives.  The implications of the varying error levels are discussed in more depth in Section \ref{Sec:ErrorDis}.  
 
\begin{figure}[ht]
     \includegraphics[width=0.47\textwidth]{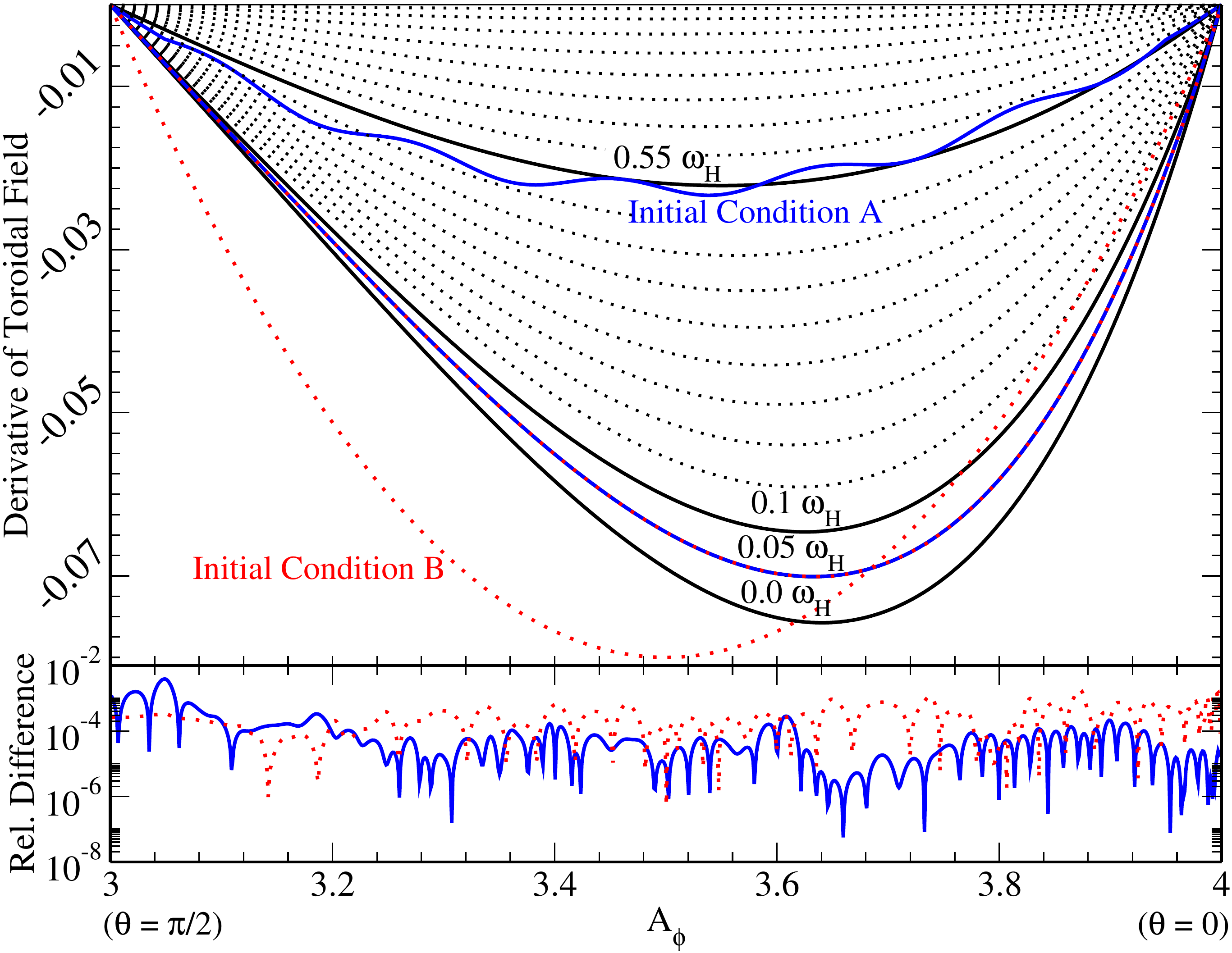}
     \caption{The derivatives of the toroidal field $d/dA_\phi (\sqrt{-g}F^{\theta r})^2$ obtained numerically for black hole spin parameter $a = 0.8m$ and various values of field line angular velocity $\Omega_\text{F}$ (in increments of $0.05 \omega_\text{H}$).  The blue and red dashed lines indicate two different initial guesses for the derivative of the toroidal field in the $\Omega_\text{F} = 0.05 \omega_\text{H}$ case.  The blue and red dotted line is the final result in both cases; they completely overlap and obscure the original result obtained using a much better initial guess.  The relative difference between the results obtained using the A and B initial conditions and the original result are shown in the bottom panel.}  
	   \label{fig:ErrorSGF}
\end{figure}

In order to reduce computation time we make a guess as to the ultimate structure of the fields to use as an initial condition.  It is therefore reasonable to ask if our guesses bias our results towards solutions that are close to that initial guess and ignore other potentially valid magnetospheres.  There are two unknown quantities that we make guesses for; the toroidal component of the vector potential, $A_\phi(r, \theta)$, and the derivative of the square of the toroidal field with respect to the vector potential as an unknown function of the vector potential, $d/dA_\phi (\sqrt{-g}F^{\theta r})^2 = F(A_\phi)$.  In order to examine how sensitive our solutions might be to changes in the initial form of these functions, we examine the magnetospheres obtained for black hole spin parameter $a = 0.8m$; the numerically obtained functions of the derivative of the toroidal field for those magnetospheres are shown in  Figure \ref{fig:ErrorSGF}. 

To asses how initial conditions might modify our results, we first coupled the numerical solution for the derivative of the toroidal field for the $\Omega_\text{F} = 0.05 \omega_\text{H}$ magnetosphere with the vector potential $A_\phi$ obtained for every other value of field line angular velocity (extrapolating outward to $r = 20m$ for the higher field line angular velocity solutions).  In every case we found that the numerical code rapidly converged to a solution essentially indistinguishable from the original $\Omega_\text{F} = 0.05 \omega_\text{H}$ solution.  There were some minor deviations at large radii, as the $1\%$ error level was achieved there last (in the original solution it was achieved last on the inner light surface), but significantly less than the deviation between adjacent $\Omega_\text{F} = 0.1 \omega_\text{H}$ and $\Omega_\text{F} = 0.0 \omega_\text{H}$ magnetospheres.  

As modifying the initial vector potential seemed to have no effect, we next examined modifying our initial guess for the derivative of the toroidal field.  Poor guesses for this function can result in differences (``kinks'') in $A_\phi$ across the inner light surface that can easily exceed the difference in $A_\phi$ between the azimuthal axis and the equatorial plane.  Diminishing inner light surface kinks takes the most computation time, so good guesses for the derivative of the toroidal field are the most critical initial condition and most likely source of any sensitivity our numerical solutions might have to initial conditions.   

For clarity we focus on two alternative initial guesses for the functional form of the derivative of the toroidal field, shown in Figure \ref{fig:ErrorSGF} as initial conditions A and B.  They are symmetric quadratics in $A_\phi$ that straddle the numerically obtained function, with the addition of periodic oscillations in Case A in an attempt to make a very poor initial guess without being overly ridiculous.  Both cases led to very large initial kinks across the inner light surface and significantly increased the computation time required to find a solution.  However the solutions obtained in both cases were again essentially indistinguishable from the original $\Omega_\text{F} = 0.05 \omega_\text{H}$ solution; the minor deviations in the derivative of the toroidal field (shown in in the bottom panel of Figure \ref{fig:ErrorSGF}) are dwarfed by the deviation between adjacent $\Omega_\text{F} = 0.1 \omega_\text{H}$ and $\Omega_\text{F} = 0.0 \omega_\text{H}$ magnetospheres.

%%%%%%%%%%%%%%%%%%%%%%%%%%%%%%%%%%%%%%%%%%%%%%%%%%%%%%%%%%%%%%%%%%%%%%%%%%%%%%%%%%%%%%%%%%%%%%%%%%%%%%%%%%%%%%%%%%%%%%%%%%%%%%%%%%%%%%
%%%%%%%%%%%%%%%%%%%%%%%%%%%%%%%%%%%%%%%%%%%%%%%%%%%%%%%%%%%%%%%%%%%%%%%%%%%%%%%%%%%%%%%%%%%%%%%%%%%%%%%%%%%%%%%%%%%%%%%%%%%%%%%%%%%%%%
%%%%%%%%%%%%%%%%%%%%%%%%%%%%%%%%%%%%%%%%%%%%%%%%%%%%%%%%%%%%%%%%%%%%%%%%%%%%%%%%%%%%%%%%%%%%%%%%%%%%%%%%%%%%%%%%%%%%%%%%%%%%%%%%%%%%%%
\Needspace{5\baselineskip}
\section{Discussion} \label{Sec:Discussion}

The most limiting assumption underlying our solutions is that of uniform field line angular velocity.  In this section we discuss that assumption in more detail, consider two extremes of the magnetospheres that we found, and briefly explore the numerical error of our solutions. 

%%%%%%%%%%%%%%%%%%%%%%%%%%%%%%%%%%%%%%%%%%%%%%%%%%%%%%%%%%%%%%%%%%%%%%%%%%%%%%%%%%%%%%%%%%%%%%%%%%%%%%%%%%%%%%%%%%%%%%%%%%%%%%%%%%%%%%
\subsection{Assumption of Uniform $\Omega_\text{F}$} \label{Sec:InitialAss}
%%%%%%%%%%%%%%%%%%%%%%%%%%%%%%%%%%%%%%%%%%%%%%%%%%%%%%%%%%%%%%%%%%%%%%%%%%%%%%%%%%%%%%%%%%%%%%%%%%%%%%%%%%%%%%%%%%%%%%%%%%%%%%%%%%%%%%

As stated in Section \ref{Sec:ModelAss}, we have assumed uniform field line angular velocities $\Omega_\text{F}$ in order to simplify the task of studying how the location of the inner Alfv\'{e}n surface might correspond to changes in the structure of energy extracting black hole magnetospheres.  This restricted us to solving for the structure of magnetospheres inside the outer light surface.  With solutions in hand we return and examine that restriction in more detail.  We begin by examining how limiting the assumption of uniform $\Omega_\text{F}$ might be in the two extreme classes of magnetospheres ($\Omega_\text{F}/\omega_\text{H} \approx 0$ and $\Omega_\text{F} \approx \omega_\text{H}$) that we found.  We then discuss the existence and uniqueness of solutions that pass smoothly through both light surfaces.  We close with a discussion of how limiting uniform $\Omega_\text{F}$ might be in interpreting our results.        

For low field line angular velocities ($\Omega_\text{F}/\omega_\text{H} \approx 0$) where the magnetic field lines bend towards the azimuthal axis the limitations imposed by solutions restricted to the interior of the outer light surface should not be a significant concern.  In that case the outer light surface can be many thousands of gravitational radii away from the horizon near the azimuthal axis and formally infinitely far away exactly on the axis.  Diminishing the kink on such a distant surface could slightly modify the structure of the magnetosphere near the horizon, but in reasonable application we would generally expect deviations from our core assumptions of stationarity, axisymmetry, a perfectly conducting force-free plasma over such an extended region to be far more significant.  Should that not be the case, we also see a smooth transition from our $\Omega_\text{F} = 0.05 \omega_\text{H}$ solutions to our $\Omega_\text{F} = 0$ solutions for which an outer light surface does not exist (formally located infinitely far away from the black hole).  We see no reason to expect pathologies in the limit $\Omega_\text{F} \rightarrow 0$, implying that our low $\Omega_\text{F}$ solutions are close to ones those that pass smoothly through both light surfaces.          

For high field line angular velocities ($\Omega_\text{F} \approx \omega_\text{H}$) where the magnetic field lines bend towards the equatorial plane we would expect to find a connection to nearby accreting matter (not necessarily a thin disk limited to the equatorial plane).  Our boundary condition of a single magnetic field line in the equatorial plane might be reasonable close to the horizon, but moving away from the horizon it would become increasingly likely to find significant deviations depending upon the specific model of nearby accreting matter that was chosen.  Finding solutions that passed smoothly through both light surfaces would therefore be somewhat ill-advised, as it would involve modifying the near horizon magnetosphere using restrictions found in unreasonably distant regions.  A more realistic magnetosphere would include a description of nearby accreting matter that included a model of plasma injection into an inflow interior to or near the separation surface.  Our solutions could be representative of such inflows, especially a thick disk or torus near the ergosphere.  A solution that passed smoothly through an outer light surface, on the other hand, would also need to include some model of plasma injection into an outflow consistent with the field line geometry obtained, which might be difficult.   

In arguing that it is possible to usefully interpret the above limits on $\Omega_\text{F}$ despite the limitations of our domain, we have ignored the question of whether or not similar solutions that pass smoothly though both light surfaces even exist.  Using essentially identical boundary conditions on the vector potential $A_\phi$, \citet{CKP2013} and \citet{NC2014} (hereafter CKP13 and NC14)  modified both the field line angular velocity and toroidal field to find solutions that pass smoothly through both light surfaces.  In doing so they found a single nearly monopolar solution with varying $\Omega_\text{F} \approx 0.5 \omega_\text{H}$ that is very similar in structure to our uniform $\Omega_\text{F} = 0.5 \omega_\text{H}$ solutions.  It is therefore natural to ask if the solutions found by CKP13 and NC14 are unique.  The exact solutions of Equation \ref{ExactSolnEqn}, while non energy extracting and largely unphysical, indicate that they are not; there are in fact infinitely many solutions fulfilling our boundary conditions on $A_\phi$ that pass smoothly through both an inner and outer light surface.  The question then becomes why we found the solutions that we did regardless of initial conditions, and why CKP13 and NC14 similarly found a single solution. 

The answer to that question lies in our numerical techniques.  We show in Appendix \ref{App:ConvProof} that the magnetofrictional method works by minimizing the energy in the electromagnetic fields.  In minimizing the kink across the inner light surface we are finding matched minimum energy solutions.  The apparent uniqueness of our results is a suggestion of the uniqueness of a minimum energy solution, not the uniqueness of our solution in full generality.  The majority of numerical codes are common in the behavior of being dissipative and seeking minimum energy states; any code that allows energy to be added at will is likely to be numerically unstable.  For example there is a full range of valid rotating Schwarzschild monopole solutions (Equation \ref{SchwMonEqn}), but it would be a very unique numerical code that would be capable of blindly finding all of them.  Most codes would always converge on the minimum energy $\Omega_0 = 0$ solution.  

For any value of black hole spin, the minimum energy magnetosphere consistent with our boundary conditions will be as close to monopolar as possible ($\Omega_\text{F} \approx 0.5 \omega_\text{H}$).  The bunching of magnetic field lines towards the axis or equator seen in our low and high $\Omega_\text{F}$ solutions requires the addition of energy to spin the magnetosphere away from that monopole.  It is therefore not surprising that CKP13 and NC14 found a single roughly monopolar solution that passes smoothly through both inner and outer light surfaces; without the explicit and very careful addition of energy to move to and maintain a rotating magnetosphere any stable code would likely converge on that solution.  Our utilization of uniform $\Omega_\text{F}$ may be interpreted as a way of exploring the structure of arbitrarily rotating magnetospheres using a dissipative numerical code, and we believe that a large fraction of our solutions could be taken to be good approximations of solutions that pass smoothly through both light surfaces.  In realistic application, however, they should largely be taken to be representative of inflow solutions.  Outflow solutions would require the additional description of plasma injection mechanisms, and even if our solutions passed smoothly through an outer light surface there is no guarantee that plasma parameters would be either continuous or conserved across an extended plasma injection region (as implied by single solutions that pass smoothly through both light surfaces).  
    
We have argued that our limited domain might not prevent useful interpretations	and that our solutions might approximate solutions that pass smoothly through both light surfaces (even if such solutions might be of dubious value).  It could still be asked how representative our solutions might be of energy extracting black hole magnetospheres, as completely uniform $\Omega_\text{F}$ magnetospheres are unlikely to exist.  We cannot fully answer that question, but we emphasize that exploring uniform $\Omega_\text{F}$ magnetospheres is not the goal of this work.  The goal of this work is to explore the effects of inner Alfv\'{e}n surface location.  Uniform $\Omega_\text{F}$ in the force-free limit is merely a useful tool to do so, both in its obvious simplicity as well as in allowing us to know exactly where the inner light surface will be prior to solving for the structure of the magnetosphere that passes through it.  Slightly deforming the location of the Alfv\'{e}n surfaces used here will not have significant effects, and in general we expect the Alfv\'{e}n surface to be continuous, so the solutions obtained here could be merged to explore more complex scenarios.  For example, if one desired an ingoing solution with a connection to accreting matter near the equator and the launching of a jet-like structure from the horizon along the rotational axis, an appeal could be made to an Alfv\'{e}n surface that lies closer to the boundary of the ergosphere at higher latitudes and closer to the horizon near the equator.  On the horizon this would correspond to lower $\Omega_\text{F}$ for small values of $\theta$ and higher $\Omega_\text{F}$ for large values of $\theta$, compatible with the simulations conducted by \citet{MG2004}.          

%%%%%%%%%%%%%%%%%%%%%%%%%%%%%%%%%%%%%%%%%%%%%%%%%%%%%%%%%%%%%%%%%%%%%%%%%%%%%%%%%%%%%%%%%%%%%%%%%%%%%%%%%%%%%%%%%%%%%%%%%%%%%%%%%%%%%%
\subsection{Two Types of Magnetospheres}
%%%%%%%%%%%%%%%%%%%%%%%%%%%%%%%%%%%%%%%%%%%%%%%%%%%%%%%%%%%%%%%%%%%%%%%%%%%%%%%%%%%%%%%%%%%%%%%%%%%%%%%%%%%%%%%%%%%%%%%%%%%%%%%%%%%%%%

Our solutions indicate two extreme methods for black hole energy to be extracted and transmitted to distant observers.  For low field line angular velocities ($\Omega_\text{F} / \omega_\text{H} \approx 0$) jet-like structures naturally form to transmit extracted energy directly from the horizon to distant observers along the azimuthal axis.  For high field line angular velocities ($\Omega_\text{F} \approx \omega_\text{H}$) we find magnetospheres that are consistent with the transmission of energy from the horizon to nearby accreting matter, which in turn might transmit the extracted energy to distant observers.  In this section we discuss potential consequences of those two types of magnetospheres, noting at that outset that any potential distinction between the two in realistic astrophysical scenarios is likely to be fuzzy.

If we examine the rate of energy and angular momentum extraction through Equations \ref{fitChiEqn} and \ref{fitVarphiEqn}, we see that the two types might have significant time-dependent distinctions due to their differing rates of angular momentum extraction and concurrent black hole spindown.  Therefore transient high energy phenomena powered by black hole energy extraction might also have two distinct signatures.  For specificity, consider a gamma-ray burst (GRB).  Black hole energy extraction is a candidate for the central engine of a GRB (e.g. \citet{GehrelsMeszaros2012}), and the exponential decay associated with black hole energy extraction might be compatible with some GRB observations (e.g. \citet{NathanailStrantzalisContopoulos2016}).  We do not demand a specific type or model of GRB, however.  We only require a transient high energy event that might be powered by black hole energy extraction, and ``GRB'' is a convenient specific stand-in for``transient high energy event'' even if our treatment is too crude to completely describe or differentiate between realistic GRBs of any specific type.  

\begin{figure}[ht]
     \includegraphics[width=0.47\textwidth]{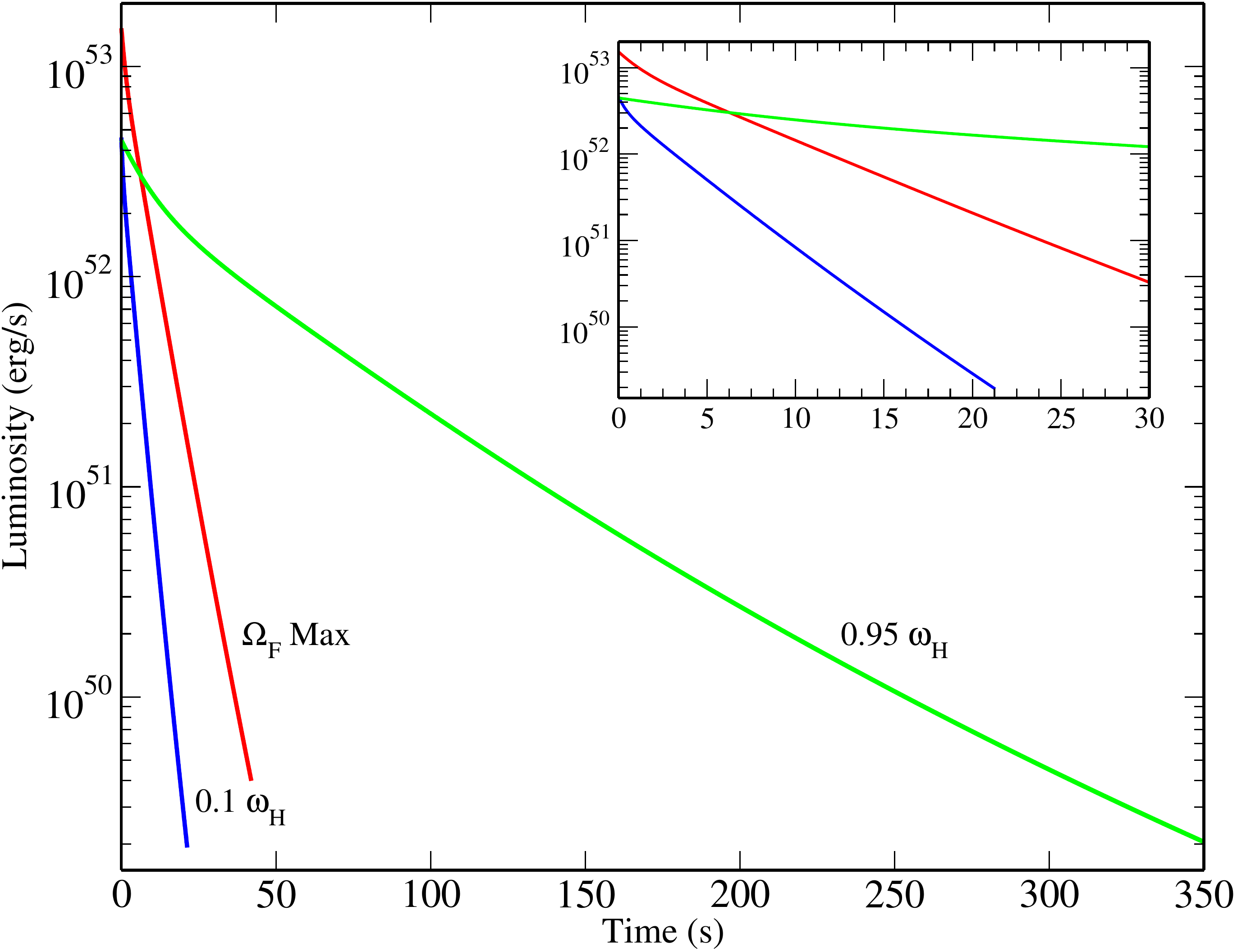}
     \caption{Rate of black hole energy extraction for constant field line angular velocities $\Omega_\text{F} = 0.1 \omega_\text{H}$, $\Omega_\text{F} = 0.95 \omega_\text{H}$, and the field line angular velocity corresponding to the maximum rate of energy extraction for a black hole with initial spin $a = 0.95m$.  The curves for $\Omega_\text{F} = 0.1 \omega_\text{H}$ and $\Omega_\text{F} = \Omega_\text{F Max}$ are terminated when the black hole's spin falls below $a=0.05m$; it takes around 520 seconds for the $\Omega_\text{F} = 0.95 \omega_\text{H}$ case to reach that level.}  
	   \label{fig:10SLum}
\end{figure}

If we take $B_x = 10^{16} \text{G}$ and $r_x = 3$ as constants in Equations \ref{P2Eqn} and \ref{Eqn:LwithBx} (i.e. a magnetic field strength of roughly $10^{16} \text{G}$ just outside the ergosphere), assume that $a = 0.95m$ and $m = 4 M_\odot$ at time $t=0$, and for simplicity insist upon a state of suspended accretion (such as a magnetically arrested disk in a low accretion state), we find the rates of black hole energy extraction plotted in Figure \ref{fig:10SLum}.      
        
We immediately note that all three cases exhibit roughly exponential decay, but the $\Omega_\text{F} = 0.95 \omega_\text{H}$ case decays relatively slowly.  After roughly 20-30 seconds both the $\Omega_\text{F} = 0.1 \omega_\text{H}$ and $\Omega_\text{F Max}$ luminosities have decreased by factors of $\sim 10^3$ due to the black hole spin dropping below $a \approx 0.05 m$.  The rate of angular momentum extraction for the $0.95 \omega_\text{H}$ case is far lower, however, and after 30 seconds its luminosity has not significantly decreased.  Note that in this model larger black hole masses would take less time to spin down due to the increase in magnetic field strength implied by $r_x$ corresponding to a larger dimensional radius.  For example, a black hole with twice the mass but otherwise identical parameters would be spun down in around half the time.  Also note that the scaling on the axes is somewhat irrelevant; they were chosen for crude correspondence to a GRB, but a wide range of systems with different parameters, luminosities, and timescales exhibit qualitatively identical curves. 

It takes over 350 seconds for the $0.95 \omega_\text{H}$ luminosity to drop to the same level that the $0.1 \omega_\text{H}$ luminosity fell to in around 20 seconds.  We therefore expect that GRBs with magnetic field lines that directly connect the horizon to distant observers might be at least 10-20 times shorter than GRBs with magnetic field lines connecting to a disk or other matter before connecting to distant observers.  The addition of more realistic effects such as accretion actively spinning up the black hole would obviously complicate matters, and as noted at the outset the boundary between both types of magnetospheres is unlikely to be very distinct.  However, if GRBs (again emphasizing the limited sense in which we are using the term) are powered by black hole energy extraction it would not be unreasonable to expect two broad classes with differing characteristic durations as well as different radiation signatures; the longer lived $0.95 \omega_\text{H}$ type might also exhibit a greater degree of thermalization due to the deposition of the extracted energy into inflowing matter before that energy is transmitted to distant observers. 

The analysis presented here is too crude to reasonably claim that distinctions between GRBs can arise due to differences in the location of their inner Alfv\'{e}n surface, encapsulated here in differences in field line angular velocity.  A more reasonable model is beyond the scope of our present work, but the possibility that such a model could allow finer classification of some GRBs or other transient high energy phenomena via correlation of timescale and degree of thermalization is interesting.   

%%%%%%%%%%%%%%%%%%%%%%%%%%%%%%%%%%%%%%%%%%%%%%%%%%%%%%%%%%%%%%%%%%%%%%%%%%%%%%%%%%%%%%%%%%%%%%%%%%%%%%%%%%%%%%%%%%%%%%%%%%%%%%%%%%%%%%
\Needspace{5\baselineskip}
\subsection{Numerical Error Analysis} \label{Sec:ErrorDis}
%%%%%%%%%%%%%%%%%%%%%%%%%%%%%%%%%%%%%%%%%%%%%%%%%%%%%%%%%%%%%%%%%%%%%%%%%%%%%%%%%%%%%%%%%%%%%%%%%%%%%%%%%%%%%%%%%%%%%%%%%%%%%%%%%%%%%%

In this section we discuss how reliable the numerical calculations underlying our solutions might be.  

With the possible exception of the kink at the inner light surface, we find that our numerical solutions are consistent with the precision of their numerical grid.  For the case of $a = 0.3m$ and $\Omega_\text{F} = 0.5 \omega_\text{H}$, the error of our solution is generally around 100 times smaller than the error in any of the perturbed monopole solutions that are considered to be good approximations at low spin.  While the numerical error involved in implementing the exact \textit{HS3} solution (arising mostly from taking numerical derivatives) can be another 100 times smaller still, there are large regions in which the numerical solution is actually ``better'' than the \textit{HS3} solution and likely exceeds the precision justified by our numerical grid.

Along the inner light surface the error in our numerical solution is generally a few times smaller than the third-order perturbed monopole solution, and exceeds the first-order solution by a factor of 10-100.  Again the \textit{HS3} solution generally exceeds the numerical solution by another factor of 10-100.  However significantly better solutions would be difficult to obtain.  At the $1\%$ level the ``kink'' (change in $A_\phi$) in our solution across the inner light surface in the radial direction is around $10^{-6}$, much smaller than the required linear change in $A_\phi$ of $5\times10^{-3}$ in the $\theta$ direction implied by our usage of 200 grid squares between the azimuthal axis (where $A_\phi = 4$) and the equator (where $A_\phi = 3$).  This means that it becomes very difficult to accurately quantify the kink, much less diminish it further.  

Some experimentation reveals that a much higher error level of around 10\% would not have changed our results in any significant way.  This is consistent with the error of the perturbed solutions, which should be accurate enough for most purposes and have similarly sized errors.  Reducing the error of our solutions to the 1\% level increased the required computation time significantly, and achieves a precision that almost certainly far exceeds what our assumptions allow for in any reasonable application.  However reducing the error to the 1\% level allows us to confidently state that any deficiencies in our results rest within our assumptions and not in our numerical calculations.

Due to the importance of selecting appropriate initial conditions in diminishing computation time, it might be asked if our initial guesses in any way affect or drive our results.  As shown in Figure \ref{fig:ErrorSGF}, even with somewhat ridiculous choices of initial condition we ultimately arrive at effectively identical solutions that are appreciably different from neighboring solutions.  We cannot (and do not) claim that this implies that our solutions are unique in general; it is possible that there are many solutions that satisfy our boundary conditions and assumptions.  The only potentially persuasive uniqueness in our solutions lies in unique minimum energy solutions that are matched across the inner light surface, as a consequence of the magnetofrictional method being an energy minimizing algorithm (shown in Appendix \ref{App:ConvProof}).  In developing our numerical code we conducted many more trials and experiments than are shown here, and we never observed any indication that it might be possible to converge on two different solutions.  Therefore while we cannot prove that our solutions are unique matched minimum energy solutions consistent with our boundary conditions and assumptions, we do believe that it is a reasonable assumption to make.               
             
%%%%%%%%%%%%%%%%%%%%%%%%%%%%%%%%%%%%%%%%%%%%%%%%%%%%%%%%%%%%%%%%%%%%%%%%%%%%%%%%%%%%%%%%%%%%%%%%%%%%%%%%%%%%%%%%%%%%%%%%%%%%%%%%%%%%%%
%%%%%%%%%%%%%%%%%%%%%%%%%%%%%%%%%%%%%%%%%%%%%%%%%%%%%%%%%%%%%%%%%%%%%%%%%%%%%%%%%%%%%%%%%%%%%%%%%%%%%%%%%%%%%%%%%%%%%%%%%%%%%%%%%%%%%%
%%%%%%%%%%%%%%%%%%%%%%%%%%%%%%%%%%%%%%%%%%%%%%%%%%%%%%%%%%%%%%%%%%%%%%%%%%%%%%%%%%%%%%%%%%%%%%%%%%%%%%%%%%%%%%%%%%%%%%%%%%%%%%%%%%%%%%

\section{Conclusions}

The defining feature of energy extracting black hole magnetospheres is the location of their inner Alfv\'{e}n surface, which coincides with the inner light surface in the force-free limit.  Despite its importance, no comprehensive studies of the effects of modifying the location of the inner Alfv\'{e}n surface have been accomplished.  We have begun addressing that deficiency by studying how simple force-free magnetospheres can be modified as a function of uniform field line angular velocity, which together with black hole spin determines the location of the inner light surface.  In order to do so we extended the Newtonian magnetofrictional method for computing force-free magnetospheres into the general relativistic regime, which allowed us to efficiently calculate hundreds of energy extracting black hole magnetospheres as functions of inner Alfv\'{e}n surface location.  In so doing we found that inner Alfv\'{e}n surfaces near the horizon cause extracted energy to flow towards the equatorial plane, while inner Alfv\'{e}n surfaces near the boundary of the ergosphere cause extracted energy to flow outwards via jet-like structures aligned with the azimuthal axis.  Applied to transient high energy phenomena, those magnetospheres imply two timescales that might differ by a factor of 20 or more.  This suggests that two classes of transient high energy phenomena might exist; shorter ones that directly connect the horizon to distant observers, and longer ones that have a disk or other significant matter separating the energy flow from the horizon to distant observers that potentially thermalizes the extracted energy, creating a different radiation signature.     

%%%%%%%%%%%%%%%%%%%%%%%%%%%%%%%%%%%%%%%%%%%%%%%%%%%%%%%%%%%%%%%%%%%%%%%%%%%%%%%%%%%%%%%%%%%%%%%%%%%%%%%%%%%%%%%%%%%%%%%%%%%%%%%%%%%%%%
%%%%%%%%%%%%%%%%%%%%%%%%%%%%%%%%%%%%%%%%%%%%%%%%%%%%%%%%%%%%%%%%%%%%%%%%%%%%%%%%%%%%%%%%%%%%%%%%%%%%%%%%%%%%%%%%%%%%%%%%%%%%%%%%%%%%%%
%%%%%%%%%%%%%%%%%%%%%%%%%%%%%%%%%%%%%%%%%%%%%%%%%%%%%%%%%%%%%%%%%%%%%%%%%%%%%%%%%%%%%%%%%%%%%%%%%%%%%%%%%%%%%%%%%%%%%%%%%%%%%%%%%%%%%%

\section*{Acknowledgments}

K.T. gratefully acknowledges useful conversations with Dana Longcope.  M.T. was supported by JSPS KAKENHI Grant Number 24540268.

%%%%%%%%%%%%%%%%%%%%%%%%%%%%%%%%%%%%%%%%%%%%%%%%%%%%%%%%%%%%%%%%%%%%%%%%%%%%%%%%%%%%%%%%%%%%%%%%%%%%%%%%%%%%%%%%%%%%%%%%%%%%%%%%%%%%%%
%%%%%%%%%%%%%%%%%%%%%%%%%%%%%%%%%%%%%%%%%%%%%%%%%%%%%%%%%%%%%%%%%%%%%%%%%%%%%%%%%%%%%%%%%%%%%%%%%%%%%%%%%%%%%%%%%%%%%%%%%%%%%%%%%%%%%%
%%%%%%%%%%%%%%%%%%%%%%%%%%%%%%%%%%%%%%%%%%%%%%%%%%%%%%%%%%%%%%%%%%%%%%%%%%%%%%%%%%%%%%%%%%%%%%%%%%%%%%%%%%%%%%%%%%%%%%%%%%%%%%%%%%%%%%

\bibliography{paper-prd}

\begin{thebibliography}{25}%
\makeatletter
\providecommand \@ifxundefined [1]{%
 \@ifx{#1\undefined}
}%
\providecommand \@ifnum [1]{%
 \ifnum #1\expandafter \@firstoftwo
 \else \expandafter \@secondoftwo
 \fi
}%
\providecommand \@ifx [1]{%
 \ifx #1\expandafter \@firstoftwo
 \else \expandafter \@secondoftwo
 \fi
}%
\providecommand \natexlab [1]{#1}%
\providecommand \enquote  [1]{``#1''}%
\providecommand \bibnamefont  [1]{#1}%
\providecommand \bibfnamefont [1]{#1}%
\providecommand \citenamefont [1]{#1}%
\providecommand \href@noop [0]{\@secondoftwo}%
\providecommand \href [0]{\begingroup \@sanitize@url \@href}%
\providecommand \@href[1]{\@@startlink{#1}\@@href}%
\providecommand \@@href[1]{\endgroup#1\@@endlink}%
\providecommand \@sanitize@url [0]{\catcode `\\12\catcode `\$12\catcode
  `\&12\catcode `\#12\catcode `\^12\catcode `\_12\catcode `\%12\relax}%
\providecommand \@@startlink[1]{}%
\providecommand \@@endlink[0]{}%
\providecommand \url  [0]{\begingroup\@sanitize@url \@url }%
\providecommand \@url [1]{\endgroup\@href {#1}{\urlprefix }}%
\providecommand \urlprefix  [0]{URL }%
\providecommand \Eprint [0]{\href }%
\providecommand \doibase [0]{http://dx.doi.org/}%
\providecommand \selectlanguage [0]{\@gobble}%
\providecommand \bibinfo  [0]{\@secondoftwo}%
\providecommand \bibfield  [0]{\@secondoftwo}%
\providecommand \translation [1]{[#1]}%
\providecommand \BibitemOpen [0]{}%
\providecommand \bibitemStop [0]{}%
\providecommand \bibitemNoStop [0]{.\EOS\space}%
\providecommand \EOS [0]{\spacefactor3000\relax}%
\providecommand \BibitemShut  [1]{\csname bibitem#1\endcsname}%
\let\auto@bib@innerbib\@empty
%</preamble>
\bibitem [{\citenamefont {{Blandford}}\ and\ \citenamefont
  {{Znajek}}(1977)}]{BZ77}%
  \BibitemOpen
  \bibfield  {author} {\bibinfo {author} {\bibfnamefont {R.~D.}\ \bibnamefont
  {{Blandford}}}\ and\ \bibinfo {author} {\bibfnamefont {R.~L.}\ \bibnamefont
  {{Znajek}}},\ }\href {\doibase 10.1093/mnras/179.3.433} {\bibfield  {journal}
  {\bibinfo  {journal} {Mon. Not. R. Astron. Soc.}\ }\textbf {\bibinfo {volume}
  {179}},\ \bibinfo {pages} {433} (\bibinfo {year} {1977})}\BibitemShut
  {NoStop}%
\bibitem [{\citenamefont {{Lasota}}\ \emph {et~al.}(2014)\citenamefont
  {{Lasota}}, \citenamefont {{Gourgoulhon}}, \citenamefont {{Abramowicz}},
  \citenamefont {{Tchekhovskoy}},\ and\ \citenamefont {{Narayan}}}]{LGATN2014}%
  \BibitemOpen
  \bibfield  {author} {\bibinfo {author} {\bibfnamefont {J.-P.}\ \bibnamefont
  {{Lasota}}}, \bibinfo {author} {\bibfnamefont {E.}~\bibnamefont
  {{Gourgoulhon}}}, \bibinfo {author} {\bibfnamefont {M.}~\bibnamefont
  {{Abramowicz}}}, \bibinfo {author} {\bibfnamefont {A.}~\bibnamefont
  {{Tchekhovskoy}}}, \ and\ \bibinfo {author} {\bibfnamefont {R.}~\bibnamefont
  {{Narayan}}},\ }\href {\doibase 10.1103/PhysRevD.89.024041} {\bibfield
  {journal} {\bibinfo  {journal} {\prd}\ }\textbf {\bibinfo {volume} {89}},\
  \bibinfo {eid} {024041} (\bibinfo {year} {2014})},\ \Eprint
  {http://arxiv.org/abs/1310.7499} {arXiv:1310.7499 [gr-qc]} \BibitemShut
  {NoStop}%
\bibitem [{\citenamefont {{Komissarov}}(2004)}]{Komissarov2004}%
  \BibitemOpen
  \bibfield  {author} {\bibinfo {author} {\bibfnamefont {S.~S.}\ \bibnamefont
  {{Komissarov}}},\ }\href {\doibase 10.1111/j.1365-2966.2004.07598.x}
  {\bibfield  {journal} {\bibinfo  {journal} {Mon. Not. R. Astron. Soc.}\
  }\textbf {\bibinfo {volume} {350}},\ \bibinfo {pages} {427} (\bibinfo {year}
  {2004})}\BibitemShut {NoStop}%
\bibitem [{\citenamefont {{Takahashi}}\ \emph {et~al.}(1990)\citenamefont
  {{Takahashi}}, \citenamefont {{Nitta}}, \citenamefont {{Tatematsu}},\ and\
  \citenamefont {{Tomimatsu}}}]{TNTT90}%
  \BibitemOpen
  \bibfield  {author} {\bibinfo {author} {\bibfnamefont {M.}~\bibnamefont
  {{Takahashi}}}, \bibinfo {author} {\bibfnamefont {S.}~\bibnamefont
  {{Nitta}}}, \bibinfo {author} {\bibfnamefont {Y.}~\bibnamefont
  {{Tatematsu}}}, \ and\ \bibinfo {author} {\bibfnamefont {A.}~\bibnamefont
  {{Tomimatsu}}},\ }\href {\doibase 10.1086/169331} {\bibfield  {journal}
  {\bibinfo  {journal} {\apj}\ }\textbf {\bibinfo {volume} {363}},\ \bibinfo
  {pages} {206} (\bibinfo {year} {1990})}\BibitemShut {NoStop}%
\bibitem [{\citenamefont {{Nitta}}\ \emph {et~al.}(1991)\citenamefont
  {{Nitta}}, \citenamefont {{Takahashi}},\ and\ \citenamefont
  {{Tomimatsu}}}]{Nitta1991}%
  \BibitemOpen
  \bibfield  {author} {\bibinfo {author} {\bibfnamefont {S.-Y.}\ \bibnamefont
  {{Nitta}}}, \bibinfo {author} {\bibfnamefont {M.}~\bibnamefont
  {{Takahashi}}}, \ and\ \bibinfo {author} {\bibfnamefont {A.}~\bibnamefont
  {{Tomimatsu}}},\ }\href {\doibase 10.1103/PhysRevD.44.2295} {\bibfield
  {journal} {\bibinfo  {journal} {\prd}\ }\textbf {\bibinfo {volume} {44}},\
  \bibinfo {pages} {2295} (\bibinfo {year} {1991})}\BibitemShut {NoStop}%
\bibitem [{\citenamefont {{McKinney}}\ and\ \citenamefont
  {{Gammie}}(2004)}]{MG2004}%
  \BibitemOpen
  \bibfield  {author} {\bibinfo {author} {\bibfnamefont {J.~C.}\ \bibnamefont
  {{McKinney}}}\ and\ \bibinfo {author} {\bibfnamefont {C.~F.}\ \bibnamefont
  {{Gammie}}},\ }\href {\doibase 10.1086/422244} {\bibfield  {journal}
  {\bibinfo  {journal} {\apj}\ }\textbf {\bibinfo {volume} {611}},\ \bibinfo
  {pages} {977} (\bibinfo {year} {2004})},\ \Eprint
  {http://arxiv.org/abs/astro-ph/0404512} {astro-ph/0404512} \BibitemShut
  {NoStop}%
\bibitem [{\citenamefont {{Pan}}\ and\ \citenamefont {{Yu}}(2015)}]{PanYu2015}%
  \BibitemOpen
  \bibfield  {author} {\bibinfo {author} {\bibfnamefont {Z.}~\bibnamefont
  {{Pan}}}\ and\ \bibinfo {author} {\bibfnamefont {C.}~\bibnamefont {{Yu}}},\
  }\href {\doibase 10.1088/0004-637X/812/1/57} {\bibfield  {journal} {\bibinfo
  {journal} {\apj}\ }\textbf {\bibinfo {volume} {812}},\ \bibinfo {eid} {57}
  (\bibinfo {year} {2015})},\ \Eprint {http://arxiv.org/abs/1504.04864}
  {arXiv:1504.04864 [astro-ph.HE]} \BibitemShut {NoStop}%
\bibitem [{\citenamefont {{Menon}}\ and\ \citenamefont
  {{Dermer}}(2007)}]{MenonDermer2007}%
  \BibitemOpen
  \bibfield  {author} {\bibinfo {author} {\bibfnamefont {G.}~\bibnamefont
  {{Menon}}}\ and\ \bibinfo {author} {\bibfnamefont {C.~D.}\ \bibnamefont
  {{Dermer}}},\ }\href {\doibase 10.1007/s10714-007-0418-2} {\bibfield
  {journal} {\bibinfo  {journal} {General Relativity and Gravitation}\ }\textbf
  {\bibinfo {volume} {39}},\ \bibinfo {pages} {785} (\bibinfo {year} {2007})},\
  \Eprint {http://arxiv.org/abs/astro-ph/0511661} {astro-ph/0511661}
  \BibitemShut {NoStop}%
\bibitem [{\citenamefont {{Gralla}}\ and\ \citenamefont
  {{Jacobson}}(2014)}]{GrallaJacobson2014}%
  \BibitemOpen
  \bibfield  {author} {\bibinfo {author} {\bibfnamefont {S.~E.}\ \bibnamefont
  {{Gralla}}}\ and\ \bibinfo {author} {\bibfnamefont {T.}~\bibnamefont
  {{Jacobson}}},\ }\href {\doibase 10.1093/mnras/stu1690} {\bibfield  {journal}
  {\bibinfo  {journal} {Mon. Not. R. Astron. Soc.}\ }\textbf {\bibinfo {volume}
  {445}},\ \bibinfo {pages} {2500} (\bibinfo {year} {2014})},\ \Eprint
  {http://arxiv.org/abs/1401.6159} {arXiv:1401.6159 [astro-ph.HE]} \BibitemShut
  {NoStop}%
\bibitem [{\citenamefont {{Contopoulos}}\ \emph {et~al.}(2013)\citenamefont
  {{Contopoulos}}, \citenamefont {{Kazanas}},\ and\ \citenamefont
  {{Papadopoulos}}}]{CKP2013}%
  \BibitemOpen
  \bibfield  {author} {\bibinfo {author} {\bibfnamefont {I.}~\bibnamefont
  {{Contopoulos}}}, \bibinfo {author} {\bibfnamefont {D.}~\bibnamefont
  {{Kazanas}}}, \ and\ \bibinfo {author} {\bibfnamefont {D.~B.}\ \bibnamefont
  {{Papadopoulos}}},\ }\href {\doibase 10.1088/0004-637X/765/2/113} {\bibfield
  {journal} {\bibinfo  {journal} {\apj}\ }\textbf {\bibinfo {volume} {765}},\
  \bibinfo {eid} {113} (\bibinfo {year} {2013})},\ \Eprint
  {http://arxiv.org/abs/1212.0320} {arXiv:1212.0320 [astro-ph.HE]} \BibitemShut
  {NoStop}%
\bibitem [{\citenamefont {{Nathanail}}\ and\ \citenamefont
  {{Contopoulos}}(2014)}]{NC2014}%
  \BibitemOpen
  \bibfield  {author} {\bibinfo {author} {\bibfnamefont {A.}~\bibnamefont
  {{Nathanail}}}\ and\ \bibinfo {author} {\bibfnamefont {I.}~\bibnamefont
  {{Contopoulos}}},\ }\href {\doibase 10.1088/0004-637X/788/2/186} {\bibfield
  {journal} {\bibinfo  {journal} {\apj}\ }\textbf {\bibinfo {volume} {788}},\
  \bibinfo {eid} {186} (\bibinfo {year} {2014})},\ \Eprint
  {http://arxiv.org/abs/1404.0549} {arXiv:1404.0549 [astro-ph.HE]} \BibitemShut
  {NoStop}%
\bibitem [{\citenamefont {{Thorne}}\ and\ \citenamefont
  {{MacDonald}}(1982)}]{ThorneMacdonald1982}%
  \BibitemOpen
  \bibfield  {author} {\bibinfo {author} {\bibfnamefont {K.~S.}\ \bibnamefont
  {{Thorne}}}\ and\ \bibinfo {author} {\bibfnamefont {D.}~\bibnamefont
  {{MacDonald}}},\ }\href {\doibase 10.1093/mnras/198.2.339} {\bibfield
  {journal} {\bibinfo  {journal} {Mon. Not. R. Astron. Soc.}\ }\textbf
  {\bibinfo {volume} {198}},\ \bibinfo {pages} {339} (\bibinfo {year}
  {1982})}\BibitemShut {NoStop}%
\bibitem [{\citenamefont {{Bekenstein}}\ and\ \citenamefont
  {{Oron}}(1978)}]{BekensteinOron1978}%
  \BibitemOpen
  \bibfield  {author} {\bibinfo {author} {\bibfnamefont {J.~D.}\ \bibnamefont
  {{Bekenstein}}}\ and\ \bibinfo {author} {\bibfnamefont {E.}~\bibnamefont
  {{Oron}}},\ }\href {\doibase 10.1103/PhysRevD.18.1809} {\bibfield  {journal}
  {\bibinfo  {journal} {\prd}\ }\textbf {\bibinfo {volume} {18}},\ \bibinfo
  {pages} {1809} (\bibinfo {year} {1978})}\BibitemShut {NoStop}%
\bibitem [{\citenamefont {{Camenzind}}(1986)}]{Camenzind1986a}%
  \BibitemOpen
  \bibfield  {author} {\bibinfo {author} {\bibfnamefont {M.}~\bibnamefont
  {{Camenzind}}},\ }\href@noop {} {\bibfield  {journal} {\bibinfo  {journal}
  {Astron. Astrophys.}\ }\textbf {\bibinfo {volume} {156}},\ \bibinfo {pages}
  {137} (\bibinfo {year} {1986})}\BibitemShut {NoStop}%
\bibitem [{\citenamefont {{Znajek}}(1977)}]{Znajek1977}%
  \BibitemOpen
  \bibfield  {author} {\bibinfo {author} {\bibfnamefont {R.~L.}\ \bibnamefont
  {{Znajek}}},\ }\href {\doibase 10.1093/mnras/179.3.457} {\bibfield  {journal}
  {\bibinfo  {journal} {Mon. Not. R. Astron. Soc.}\ }\textbf {\bibinfo {volume}
  {179}},\ \bibinfo {pages} {457} (\bibinfo {year} {1977})}\BibitemShut
  {NoStop}%
\bibitem [{\citenamefont {{Takahashi}}(2002)}]{Takahashi2002}%
  \BibitemOpen
  \bibfield  {author} {\bibinfo {author} {\bibfnamefont {M.}~\bibnamefont
  {{Takahashi}}},\ }\href {\doibase 10.1086/339497} {\bibfield  {journal}
  {\bibinfo  {journal} {\apj}\ }\textbf {\bibinfo {volume} {570}},\ \bibinfo
  {pages} {264} (\bibinfo {year} {2002})},\ \Eprint
  {http://arxiv.org/abs/astro-ph/0201327} {astro-ph/0201327} \BibitemShut
  {NoStop}%
\bibitem [{\citenamefont {{MacDonald}}\ and\ \citenamefont
  {{Thorne}}(1982)}]{MacDonaldThorne1982}%
  \BibitemOpen
  \bibfield  {author} {\bibinfo {author} {\bibfnamefont {D.}~\bibnamefont
  {{MacDonald}}}\ and\ \bibinfo {author} {\bibfnamefont {K.~S.}\ \bibnamefont
  {{Thorne}}},\ }\href {\doibase 10.1093/mnras/198.2.345} {\bibfield  {journal}
  {\bibinfo  {journal} {Mon. Not. R. Astron. Soc.}\ }\textbf {\bibinfo {volume}
  {198}},\ \bibinfo {pages} {345} (\bibinfo {year} {1982})}\BibitemShut
  {NoStop}%
\bibitem [{\citenamefont {{Uzdensky}}(2005)}]{Uzdensky2005}%
  \BibitemOpen
  \bibfield  {author} {\bibinfo {author} {\bibfnamefont {D.~A.}\ \bibnamefont
  {{Uzdensky}}},\ }\href {\doibase 10.1086/427180} {\bibfield  {journal}
  {\bibinfo  {journal} {\apj}\ }\textbf {\bibinfo {volume} {620}},\ \bibinfo
  {pages} {889} (\bibinfo {year} {2005})},\ \Eprint
  {http://arxiv.org/abs/astro-ph/0410715} {astro-ph/0410715} \BibitemShut
  {NoStop}%
\bibitem [{\citenamefont {{Fendt}}(1997)}]{Fendt1997}%
  \BibitemOpen
  \bibfield  {author} {\bibinfo {author} {\bibfnamefont {C.}~\bibnamefont
  {{Fendt}}},\ }\href@noop {} {\bibfield  {journal} {\bibinfo  {journal}
  {Astron. Astrophys.}\ }\textbf {\bibinfo {volume} {319}},\ \bibinfo {pages}
  {1025} (\bibinfo {year} {1997})}\BibitemShut {NoStop}%
\bibitem [{\citenamefont {{Yang}}\ \emph {et~al.}(1986)\citenamefont {{Yang}},
  \citenamefont {{Sturrock}},\ and\ \citenamefont {{Antiochos}}}]{YSA1986}%
  \BibitemOpen
  \bibfield  {author} {\bibinfo {author} {\bibfnamefont {W.~H.}\ \bibnamefont
  {{Yang}}}, \bibinfo {author} {\bibfnamefont {P.~A.}\ \bibnamefont
  {{Sturrock}}}, \ and\ \bibinfo {author} {\bibfnamefont {S.~K.}\ \bibnamefont
  {{Antiochos}}},\ }\href {\doibase 10.1086/164610} {\bibfield  {journal}
  {\bibinfo  {journal} {\apj}\ }\textbf {\bibinfo {volume} {309}},\ \bibinfo
  {pages} {383} (\bibinfo {year} {1986})}\BibitemShut {NoStop}%
\bibitem [{\citenamefont {{Hawley}}\ \emph {et~al.}(1984)\citenamefont
  {{Hawley}}, \citenamefont {{Smarr}},\ and\ \citenamefont
  {{Wilson}}}]{HSW1984}%
  \BibitemOpen
  \bibfield  {author} {\bibinfo {author} {\bibfnamefont {J.~F.}\ \bibnamefont
  {{Hawley}}}, \bibinfo {author} {\bibfnamefont {L.~L.}\ \bibnamefont
  {{Smarr}}}, \ and\ \bibinfo {author} {\bibfnamefont {J.~R.}\ \bibnamefont
  {{Wilson}}},\ }\href {\doibase 10.1086/190953} {\bibfield  {journal}
  {\bibinfo  {journal} {Astrophys. J. Suppl. Ser.}\ }\textbf {\bibinfo {volume}
  {55}},\ \bibinfo {pages} {211} (\bibinfo {year} {1984})}\BibitemShut
  {NoStop}%
\bibitem [{\citenamefont {{Contopoulos}}\ \emph {et~al.}(1999)\citenamefont
  {{Contopoulos}}, \citenamefont {{Kazanas}},\ and\ \citenamefont
  {{Fendt}}}]{Contopoulos1999}%
  \BibitemOpen
  \bibfield  {author} {\bibinfo {author} {\bibfnamefont {I.}~\bibnamefont
  {{Contopoulos}}}, \bibinfo {author} {\bibfnamefont {D.}~\bibnamefont
  {{Kazanas}}}, \ and\ \bibinfo {author} {\bibfnamefont {C.}~\bibnamefont
  {{Fendt}}},\ }\href {\doibase 10.1086/306652} {\bibfield  {journal} {\bibinfo
   {journal} {\apj}\ }\textbf {\bibinfo {volume} {511}},\ \bibinfo {pages}
  {351} (\bibinfo {year} {1999})},\ \Eprint
  {http://arxiv.org/abs/astro-ph/9903049} {astro-ph/9903049} \BibitemShut
  {NoStop}%
\bibitem [{\citenamefont {{Malanushenko}}\ \emph {et~al.}(2014)\citenamefont
  {{Malanushenko}}, \citenamefont {{Schrijver}}, \citenamefont {{DeRosa}},\
  and\ \citenamefont {{Wheatland}}}]{MSDW2014}%
  \BibitemOpen
  \bibfield  {author} {\bibinfo {author} {\bibfnamefont {A.}~\bibnamefont
  {{Malanushenko}}}, \bibinfo {author} {\bibfnamefont {C.~J.}\ \bibnamefont
  {{Schrijver}}}, \bibinfo {author} {\bibfnamefont {M.~L.}\ \bibnamefont
  {{DeRosa}}}, \ and\ \bibinfo {author} {\bibfnamefont {M.~S.}\ \bibnamefont
  {{Wheatland}}},\ }\href {\doibase 10.1088/0004-637X/783/2/102} {\bibfield
  {journal} {\bibinfo  {journal} {\apj}\ }\textbf {\bibinfo {volume} {783}},\
  \bibinfo {eid} {102} (\bibinfo {year} {2014})},\ \Eprint
  {http://arxiv.org/abs/1312.5389} {arXiv:1312.5389 [astro-ph.SR]} \BibitemShut
  {NoStop}%
\bibitem [{\citenamefont {{Gehrels}}\ and\ \citenamefont
  {{M{\'e}sz{\'a}ros}}(2012)}]{GehrelsMeszaros2012}%
  \BibitemOpen
  \bibfield  {author} {\bibinfo {author} {\bibfnamefont {N.}~\bibnamefont
  {{Gehrels}}}\ and\ \bibinfo {author} {\bibfnamefont {P.}~\bibnamefont
  {{M{\'e}sz{\'a}ros}}},\ }\href {\doibase 10.1126/science.1216793} {\bibfield
  {journal} {\bibinfo  {journal} {Science}\ }\textbf {\bibinfo {volume}
  {337}},\ \bibinfo {pages} {932} (\bibinfo {year} {2012})},\ \Eprint
  {http://arxiv.org/abs/1208.6522} {arXiv:1208.6522 [astro-ph.HE]} \BibitemShut
  {NoStop}%
\bibitem [{\citenamefont {{Nathanail}}\ \emph {et~al.}(2016)\citenamefont
  {{Nathanail}}, \citenamefont {{Strantzalis}},\ and\ \citenamefont
  {{Contopoulos}}}]{NathanailStrantzalisContopoulos2016}%
  \BibitemOpen
  \bibfield  {author} {\bibinfo {author} {\bibfnamefont {A.}~\bibnamefont
  {{Nathanail}}}, \bibinfo {author} {\bibfnamefont {A.}~\bibnamefont
  {{Strantzalis}}}, \ and\ \bibinfo {author} {\bibfnamefont {I.}~\bibnamefont
  {{Contopoulos}}},\ }\href {\doibase 10.1093/mnras/stv2558} {\bibfield
  {journal} {\bibinfo  {journal} {Mon. Not. R. Astron. Soc.}\ }\textbf
  {\bibinfo {volume} {455}},\ \bibinfo {pages} {4479} (\bibinfo {year}
  {2016})},\ \Eprint {http://arxiv.org/abs/1507.02143} {arXiv:1507.02143
  [astro-ph.HE]} \BibitemShut {NoStop}%
\end{thebibliography}%

%%%%%%%%%%%%%%%%%%%%%%%%%%%%%%%%%%%%%%%%%%%%%%%%%%%%%%%%%%%%%%%%%%%%%%%%%%%%%%%%%%%%%%%%%%%%%%%%%%%%%%%%%%%%%%%%%%%%%%%%%%%%%%%%%%%%%%
%%%%%%%%%%%%%%%%%%%%%%%%%%%%%%%%%%%%%%%%%%%%%%%%%%%%%%%%%%%%%%%%%%%%%%%%%%%%%%%%%%%%%%%%%%%%%%%%%%%%%%%%%%%%%%%%%%%%%%%%%%%%%%%%%%%%%%
%%%%%%%%%%%%%%%%%%%%%%%%%%%%%%%%%%%%%%%%%%%%%%%%%%%%%%%%%%%%%%%%%%%%%%%%%%%%%%%%%%%%%%%%%%%%%%%%%%%%%%%%%%%%%%%%%%%%%%%%%%%%%%%%%%%%%%

\begin{appendix}

%%%%%%%%%%%%%%%%%%%%%%%%%%%%%%%%%%%%%%%%%%%%%%%%%%%%%%%%%%%%%%%%%%%%%%%%%%%%%%%%%%%%%%%%%%%%%%%%%%%%%%%%%%%%%%%%%%%%%%%%%%%%%%%%%%%%%%
\section{Proof of Convergence} \label{App:ConvProof}
%%%%%%%%%%%%%%%%%%%%%%%%%%%%%%%%%%%%%%%%%%%%%%%%%%%%%%%%%%%%%%%%%%%%%%%%%%%%%%%%%%%%%%%%%%%%%%%%%%%%%%%%%%%%%%%%%%%%%%%%%%%%%%%%%%%%%%

In this section we show that the magnetofrictional method inevitably drives a given vector potential $A_\phi$ towards a force-free state.  We do this in two steps; first, we show that a force-free state is a minimum (extremum) in the electromagnetic energy of a given volume.  Second, we show that the magnetofrictional method reduces the electromagnetic energy contained in a given volume, thereby inevitably driving that volume towards a minimum energy and force-free state.  For simplicity we assume Boyer-Lindquist coordinates, that a force-free state exists, and that no pathologies develop (e.g. magnetic reconnection, if necessary, is assumed to occur via numerical diffusion).    

We define the energy $W$ contained within the fields at a given coordinate time to be:
\begin{equation}
W \equiv \int T_t{}^t \sqrt{\gamma} d^3x.
\end{equation}
Here $\sqrt{\gamma} d^3x$ is the appropriate proper volume element and $T^{\alpha \beta}$ is the electromagnetic stress energy tensor.  Differentiating with respect to coordinate time, we find:  
\begin{align} \label{EnergyDerivativeEqn}
W_{, t} &= \int T_t{}^t{}_{; t} \sqrt{\gamma} d^3x \nonumber \\
&= \int \left(T_t{}^\alpha{}_{; \alpha} - T_t{}^a{}_{; a}\right) \sqrt{\gamma} d^3x \nonumber \\
&= -\int F_{t \beta} J^\beta \sqrt{\gamma} d^3x - \oint T_t{}^A d \Sigma_A.
\end{align}
Here $d \Sigma_A$ is the appropriate directed surface element bounding the region such that the second term is a measure of the net Poynting flux through the region's boundary.  We have assumed no field pathologies, so any field line entering the volume also exits the volume.  Energy flux per unit field line is conserved, so there can be no net Poynting flux into the region along any magnetic field lines and the second term necessarily vanishes.  In a force-free configuration $F_{\alpha \beta} J^\beta = 0$ and the first term also vanishes; therefore a force-free configuration extremises the field energy $W$ in coordinate time.

The core conceit of the magnetofrictional method is that the excess momentum flux of a non force-free configuration may be converted into the coordinate velocity of a fictitious plasma; mathematically, this may be stated as:
\begin{equation}
-F_{a \beta} J^\beta = \frac{1}{\nu} v_a. 
\end{equation}         
Here the coordinate velocity $v^a$ of the plasma is defined in terms of its four velocity $u^a$ as $v^a \equiv u^a/{u^t}$.  Both the field line angular velocity $\Omega_\text{F}$ and the toroidal field $\sqrt{-g} F^{\theta r}$ are assumed to be functions of $A_\phi$ (i.e. conserved along magnetic field lines), so $v^\phi = 0$.  We now make the simplifying assumption that $\Omega_\text{F}$ vanishes (i.e. $F_{tr} = F_{t \theta} = 0$ and there is no electric field from the perspective of a distant observer); we will address the $\Omega_\text{F} \neq 0$ case later.  We then add in a fictitious electric field that allows $v^A$ to fulfill the condition of a perfectly conducting plasma, $F_{\alpha \beta} u^\beta = 0$:
\begin{align} \label{FictitiousElectricFieldEqn}
F_{t r} &= - F_{\theta r} v^\theta, \nonumber \\
F_{t \theta} &= F_{\theta r} v^r, \nonumber \\
F_{t \phi} &= - F_{r \phi} v^r - F_{\theta \phi} v^\theta.
\end{align} 
This is analagous to defining $\mathbf{E} = -\mathbf{v} \mathbf{\times} \mathbf{B}$.  From Maxwell's equations we note that $\mathbf{\nabla} \mathbf{\times} \mathbf{E} = - \partial_t \mathbf{B}$, so defining the electric field in this way should be interpreted as defining the time rate of change of the magnetic field.  Now that the condition of a perfectly conducting plasma has been met, we simplify the first integrand for the rate of change of the field energy $W_{, t}$ (Equation \ref{EnergyDerivativeEqn}) using $F_{\alpha \beta} u^\beta = 0$ to find:
\begin{align}
F_{t \beta} J^\beta &= \left(F_{bc} v^c \right) J^b \nonumber \\
&= v^c \left(F_{\beta c} J^\beta - F_{tc} J^t \right) \nonumber \\
&= \frac{1}{\nu} v^c v_c.
\end{align}  
Therefore the rate of change of the field energy for a non force-free configuration under application of the magnetofrictional method is given by:
\begin{equation}
W_{, t} = -\int \frac{1}{\nu} v^A v_A \sqrt{\gamma} d^3x.
\end{equation}
For $\nu < 0$ the right hand side is always negative (recall the signature of our metric) and using a fictitious plasma and electric field to evolve $A_\phi$ via the condition of a perfectly conducting plasma inevitably drives the field energy towards an extremum and a force-free state.

We now address the case of $\Omega_\text{F} \neq 0$, where the electric field according to a distant observer does not originally vanish.  Treating this case separately is not necessary, but makes the above somewhat more transparent and allows us to explore the meaning of the field line angular velocity.  First, consider a subvolume over which $\Omega_\text{F}$ may be considered to be a constant.  In that subvolume the vector potential $A_\alpha = (A_t, A_r, A_\theta, A_\phi)$ is given by:
\begin{equation}
A_\alpha = \left(-\Omega_\text{F} A_\phi, A_r, A_\theta, A_\phi\right).
\end{equation}   
Suppose we now make a boost to a new frame via $\phi' = \phi - \Omega_\text{F} t$.  Then $A_{\alpha'}$ is given by:
\begin{equation}
A_{\alpha'} = \left(0, A_r, A_\theta, A_\phi\right).
\end{equation}
Noting that $F_{tr} = -A_{t, r}$ and $F_{t \theta} = -A_{t, \theta}$, we see that in the new frame $F_{tr}$, $F_{t \theta}$, and $\Omega_\text{F}$ all vanish (suggestive of the fact that $\Omega_\text{F}$ may be interpreted as a measure of the rotation of magnetic field lines referenced to zero angular momentum frames).  In this frame we are free to define an electric field that will evolve the magnetic field towards a force-free configuration using the procedure outlined above.  Noting that $t = t'$, $r = r'$, $\theta = \theta'$, and $v^{A'} = v^A$ we find that $A_{\phi'}(r', \theta') = A_\phi(r, \theta)$, so $A_{\phi', t'} = -v^{A'} A_{\phi', A'} = -v^A A_{\phi, A} = A_{\phi, t}$.  Therefore application of the magnetofrictional method in the original frame inevitably moves the vector potential towards a force-free configuration for all field line angular velocities.

%%%%%%%%%%%%%%%%%%%%%%%%%%%%%%%%%%%%%%%%%%%%%%%%%%%%%%%%%%%%%%%%%%%%%%%%%%%%%%%%%%%%%%%%%%%%%%%%%%%%%%%%%%%%%%%%%%%%%%%%%%%%%%%%%%%%%%
\section{Expanded Magnetofrictional Terms} \label{App:MFExpand}
%%%%%%%%%%%%%%%%%%%%%%%%%%%%%%%%%%%%%%%%%%%%%%%%%%%%%%%%%%%%%%%%%%%%%%%%%%%%%%%%%%%%%%%%%%%%%%%%%%%%%%%%%%%%%%%%%%%%%%%%%%%%%%%%%%%%%%

In this section we expand the advection equation of the magnetofrictional method, $A_{\phi, t} = -v^A A_{\phi, A}$, in order to explicitly show the form of the equations being used.  We begin by noting that the advection equation may be rewritten as:
\begin{equation}
A_{\phi, t} = -\nu f \left(g^{rr} A_{\phi, r}^2 + g^{\theta \theta} A_{\phi, \theta}^2 \right).
\end{equation}
Here we have exploited the fact that the coordinate velocity may be rewritten as $v_A = -\nu F_{A \beta} J^\beta = \nu f A_{\phi, A}$ for a function $f(r, \theta, A_{\phi}, \Omega_\text{F}, \sqrt{-g}F^{\theta r})$.  We then weight the coefficient of friction $\nu$ by a measure of the poloidal magnetic field strength:
\begin{equation}
\nu = \nu_0 \frac{1}{\left|B_\text{p} \right|^2}.
\end{equation}
Here $\nu_0$ is a constant; it is negative so that the magnetofrictional method converges and its magnitude is selected for numerical stability.  We define the magnitude of the poloidal field $\left|B_\text{p} \right|^2 $ as:
\begin{equation}
\left|B_\text{p} \right|^2 \equiv \frac{1}{\Sigma \Delta \sin^2 \theta} \left(A_{\phi, \theta}^2 + \Delta A_{\phi, r}^2 \right). 
\end{equation}
This weighting of $\nu$ is chosen primarily for convenience, in that it prevents regions with relatively sharp gradients in $A_\phi$ (i.e. large poloidal field) from evolving too rapidly and its factor of $\Delta$ removes the coordinate singularity on the horizon.  Using this weighting, the poloidal velocities $v^r$ and $v^\theta$ are given by:
\begin{align} \label{Eqn:Vexplicit1}
v^r &= -\nu_0 \frac{f}{\Sigma \left|B_\text{p} \right|^2} \Delta A_{\phi, r}, \nonumber \\
v^\theta &= -\nu_0 \frac{f}{\Sigma \left|B_\text{p} \right|^2}   A_{\phi, \theta}.
\end{align}
The common prefactor may be expanded as:
\begin{align} \label{Eqn:Vexplicit2}
\frac{f}{\Sigma \left|B_\text{p} \right|^2} &= \frac{1}{A_{\phi, \theta}^2 + \Delta A_{\phi, r}^2} \frac{1}{4 \pi \Sigma^2} \left[ C_{B_\phi} \frac{d}{d A_\phi} \left(\sqrt{-g} F^{\theta r} \right)^2 \right. \nonumber \\
&+ \left. \vphantom{\frac{d}{d A_\phi} \left(\sqrt{-g} F^{\theta r} \right)^2} C_r A_{\phi, r} + C_{rr} A_{\phi, rr} + C_{\Omega r} \Omega_{\text{F}, r} \right. \nonumber \\
&+ \left. \vphantom{\frac{d}{d A_\phi} \left(\sqrt{-g} F^{\theta r} \right)^2} C_{\theta} A_{\phi, \theta} + C_{\theta \theta} A_{\phi, \theta \theta} + C_{\Omega \theta} \Omega_{\text{F}, \theta} \right].
\end{align}
The factor $C_{B_\phi}$ is given by:
\begin{equation}
C_{B_\phi} = -\frac{1}{2} \Sigma^2. 
\end{equation}
$C_r$ is given by:
\begin{align}
C_r &= -\Sigma \Delta \alpha_{, r}  \nonumber \\
&= -\frac{2 m \Delta }{\Sigma} \left( r^2 - a^2 \cos^2 \theta \right) \nonumber \\
&+ \frac{4am \Delta}{\Sigma} \left( r^2 - a^2 \cos^2 \theta \right)\sin^2 \theta \cdot \Omega_\text{F} \nonumber \\
&+ \frac{2\Delta}{\Sigma} \left(r^5 + 2a^2r^3 \cos^2 \theta - a^2 m r^2 \sin^2 \theta \right. \nonumber \\
&+ \left. \vphantom{r^5} a^4 r \cos^4 \theta + a^4 m \cos^2 \theta \sin^2 \theta \right) \sin^2 \theta \cdot \Omega_\text{F}^2  \nonumber \\
&-4amr \Delta \sin^2 \theta \cdot \Omega_{\text{F}, r} \nonumber \\
&+ \Delta \left[r^4 + a^2 r^2 \left(2 - \sin^2 \theta \right) + 2a^2 m r \sin^2 \theta \right. \nonumber \\
&+ \left. \vphantom{r^4} a^4 \cos^2 \theta \right] \sin^2 \theta \cdot 2 \Omega_\text{F} \Omega_{\text{F}, r}.
\end{align}
$C_{rr}$ is given by:
\begin{align}
C_{rr} &= -\Sigma \Delta \alpha \nonumber \\
&= -\Delta \left(r^2 - 2mr + a^2 \cos^2 \theta\right) \nonumber \\
&- 4amr \Delta \sin^2 \theta \cdot \Omega_\text{F}  \nonumber \\
&+ \Delta \left[r^4 + a^2 r^2 \left(2 - \sin^2 \theta \right) \right. \nonumber \\
&+ \left. \vphantom{r^4} 2a^2 m r \sin^2 \theta + a^4 \cos^2 \theta \right] \sin^2 \theta \cdot \Omega_\text{F}^2.  
\end{align}
$C_\theta$ is given by:
\begin{align}
C_\theta &= - \Sigma \sin \theta \left[\frac{\alpha}{\sin \theta} \right]_{, \theta} \nonumber \\
&= \frac{1}{\Sigma} \left[r^4 - 2mr^3 + 2a^2 r^2 \cos^2 \theta \right. \nonumber \\
&- \left. \vphantom{r^4} 2 a^2 m r \left(1 - 3 \sin^2 \theta \right) + a^4 \cos^4 \theta \right] \frac{\cos \theta}{\sin \theta} \nonumber \\
&- \frac{4amr}{\Sigma} \left[r^2 + a^2 \left(1 + \sin^2 \theta \right)\right] \sin \theta \cos \theta \cdot \Omega_\text{F}  \nonumber \\
&+ \frac{1}{\Sigma} \left[r^6 + a^2 r^4 \left(3 - 2 \sin^2 \theta \right) + 6a^2mr^3 \sin^2 \theta \right. \nonumber \\
&+ \left. \vphantom{r^6} a^4 r^2 \left(3 - \sin^2\theta \right) \cos^2 \theta + a^4 m r \left(6 - 2 \sin^2 \theta \right) \sin^2 \theta \right. \nonumber \\
&+ \left. \vphantom{r^6} a^6 \cos^4 \theta \right] \sin \theta \cos \theta \cdot \Omega_\text{F}^2 \nonumber \\
&- 4amr \sin^2 \theta \cdot \Omega_{\text{F}, \theta} \nonumber \\
&+ \left[r^4 + a^2 r^2 \left(2 - \sin^2 \theta \right) + 2a^2 m r \sin^2 \theta \right. \nonumber \\ 
&+ \left. \vphantom{r^4} a^4 \cos^2 \theta \right] \sin^2 \theta \cdot 2 \Omega_\text{F} \Omega_{\text{F}, \theta}.
\end{align}
$C_{\theta \theta}$ is given by:
\begin{align}
C_{\theta \theta} &= - \Sigma \alpha \nonumber \\
&= -\left(r^2 - 2mr + a^2 \cos^2 \theta \right) \nonumber \\
&- 4amr \sin^2 \theta \cdot \Omega_\text{F} \nonumber \\
&+ \left[r^4 + a^2 r^2 \left(2 - \sin^2 \theta \right) + 2a^2 m r \sin^2 \theta \right. \nonumber \\
&+ \left. \vphantom{r^4} a^4 \cos^2 \theta \right] \sin^2 \theta \cdot \Omega_\text{F}^2.   
\end{align}
$C_{\Omega r}$ is given by:
\begin{align}
C_{\Omega r} &= \Sigma \Delta  G_\phi A_{\phi, r} = \Sigma \Delta \left( g_{t \phi} + g_{\phi \phi} \Omega_\text{F} \right) A_{\phi, r} \nonumber \\
&= 2 a m r \Delta A_{\phi, r} \sin^2 \theta \nonumber \\
&- \Delta \left[r^4 + a^2 r^2 \left(2 - \sin^2 \theta \right) + 2a^2 m r \sin^2 \theta \right. \nonumber \\
&+ \left. \vphantom{r^4} a^4 \cos^2 \theta \right] \sin^2 \theta \cdot \Omega_\text{F} A_{\phi, r}.
\end{align}
$C_{\Omega \theta}$ is given by:
\begin{align}
C_{\Omega \theta}  &= \Sigma G_\phi A_{\phi, \theta} = \Sigma \left( g_{t \phi} + g_{\phi \phi} \Omega_\text{F} \right) A_{\phi, \theta} \nonumber \\
&= 2 a m r A_{\phi, \theta} \sin^2 \theta \nonumber \\
&- \left[r^4 + a^2 r^2 \left(2 - \sin^2 \theta \right) + 2a^2 m r \sin^2 \theta \right. \nonumber \\
&+ \left. \vphantom{r^4} a^4 \cos^2 \theta \right] \sin^2 \theta \cdot \Omega_\text{F} A_{\phi, \theta}.
\end{align}
In this form it is obvious that there are no divergences on the horizon; as mentioned above this is a consequence of scaling $\nu$ by the magnitude of the poloidal field.  There is however a $\sin \theta$ divergence in $C_\theta$; this is enforcing $A_{\phi, \theta} = 0$ on the axis as a consequence axisymmetry.  Note also that inside the horizon $C_{rr}$ and $C_{\theta \theta}$ differ in sign; this leads to the magnetofrictional method becoming intrinsically anti-diffusive and numerically unstable there.  As mentioned in the main text, outside the light surfaces an overall factor of $-1$ is added to the above in order to maintain numerical stability, as otherwise the sign of $\alpha$ in the $C_{rr}$ and $C_{\theta \theta}$ terms leads to anti-diffusive numerical instabilities.  We finally note that all derivatives of $A_\phi$ used to evaluate $v^A$ may be taken using central finite differencing.  A one-sided derivative appropriate to upwind differencing is only taken to evolve the $A_{\phi, A}$ derivative in $A_{\phi, t} = -v^A A_{\phi, A}$.

The derivative of the toroidal field (and field line angular velocity, in cases where it is not uniform) is taken to be an unknown function of $A_\phi$.  Practically we implement it using a lookup table that is modified during runtime to diminish any kinks that develop on the inner light surface. 

%%%%%%%%%%%%%%%%%%%%%%%%%%%%%%%%%%%%%%%%%%%%%%%%%%%%%%%%%%%%%%%%%%%%%%%%%%%%%%%%%%%%%%%%%%%%%%%%%%%%%%%%%%%%%%%%%%%%%%%%%%%%%%%%%%%%%%
\section{Expanded Percent Terms} \label{PercentAppendix}
%%%%%%%%%%%%%%%%%%%%%%%%%%%%%%%%%%%%%%%%%%%%%%%%%%%%%%%%%%%%%%%%%%%%%%%%%%%%%%%%%%%%%%%%%%%%%%%%%%%%%%%%%%%%%%%%%%%%%%%%%%%%%%%%%%%%%%

In this section we detail our measure of how force-free a given configuration is.  Note that the force-free equations may be written as:
\begin{equation}
-F_{\alpha \beta} J^\beta = X_\alpha.
\end{equation}
For a configuration to be force-free we must have $X_\alpha = 0$.  If the toroidal field $\sqrt{-g} F^{\theta r}$ is conserved along magnetic field lines, as it must be when implemented as a function of $A_\phi$, then to within numerical error $X_t = X_\phi = 0.$  The poloidal components of the momentum flux may be written as:
\begin{equation}
X_A = f \cdot A_{\phi, A}. 
\end{equation}
Here $f$ is a function of $r$, $\theta$, $\Omega_\text{F}$, $\sqrt{-g} F^{\theta r}$, and $A_\phi$; setting $f=0$ yields the transfield equation.  To measure how force-free a given solution is, we must measure how close to zero $f$ is; explicitly, $f$ is given by:
\begin{equation}
f = -\frac{1}{4 \pi \Sigma \Delta \sin^3 \theta} \left[D_1 + D_2 + D_3 + D_4 + D_5 + D_6 + D_7 \right].
\end{equation}
The functions $D_i$ are given by:
\begin{align}
D_1 &= \frac{1}{2} \Sigma \sin \theta \frac{d}{d A_\phi} \left(\sqrt{-g} F^{\theta r} \right)^2 \nonumber \\
&= -\frac{\sin \theta}{\Sigma} \cdot C_{B_\phi} \frac{d}{d A_\phi} \left(\sqrt{-g} F^{\theta r} \right)^2 \nonumber \\  
D_2 &= \Delta \alpha_{, r} \sin \theta  A_{\phi, r} = -\frac{\sin \theta}{\Sigma} \cdot C_r A_{\phi, r}  \nonumber \\ 
D_3 &= \Delta \alpha \sin \theta A_{\phi, rr}  = -\frac{\sin \theta}{\Sigma} \cdot C_{rr} A_{\phi, rr}  \nonumber \\ 
D_4 &= \sin^2 \theta \left(\frac{\alpha}{\sin \theta} \right)_{, \theta} A_{\phi, \theta} = -\frac{\sin \theta}{\Sigma} \cdot C_\theta A_{\phi, \theta}  \nonumber \\ 
D_5 &= \alpha \sin \theta A_{\phi, \theta \theta} = -\frac{\sin \theta}{\Sigma} \cdot C_{\theta \theta} A_{\phi, \theta \theta}  \nonumber \\ 
D_6 &= -\Delta G_\phi \sin \theta A_{\phi, r} \Omega_{\text{F}, r} = -\frac{\sin \theta}{\Sigma} \cdot C_{\Omega r} \Omega_{\text{F}, r}  \nonumber \\ 
D_7 &= -G_\phi \sin \theta A_{\phi, \theta} \Omega_{\text{F}, \theta} = -\frac{\sin \theta}{\Sigma} \cdot C_{\Omega \theta} \Omega_{\text{F}, \theta} 
\end{align}
To establish a measure of force-freeness, we take the sum $\sum_i D_i$ and compare it to the (absolute) maximum $D_i$ term, and insist that the ratio fall below a given threshhold:
\begin{equation}
\epsilon = \frac{\sum_i D_i}{\text{Max}\left( \left|D_i\right| \right)}.
\end{equation} 
In this work we insisted that $\epsilon < 1 \%$ over the entire domain.  In practice our algorithm yields $\epsilon \ll 1\%$ over most of the domain with $\epsilon \approx 1\%$ only on some segments of the inner light surface.

\end{appendix}

%%%%%%%%%%%%%%%%%%%%%%%%%%%%%%%%%%%%%%%%%%%%%%%%%%%%%%%%%%%%%%%%%%%%%%%%%%%%%%%%%%%%%%%%%%%%%%%%%%%%%%%%%%%%%%%%%%%%%%%%%%%%%%%%%%%%%%%%%%%%%%%%%%%%%%%%%%%%%%%%%%%%%%%%%%%%%%%%%%%%%%%%%%%%%%%%%%%%%%%%%%%%%%%%%%%%%%%%%%%%%%%%%%%%%%%%%%%%%%%%%%%%%%%%%%%%%%%%%%%%%%%%%%%%%%%%%%%%%%%%%%%%%%%%%%%%%%%%%%%%%%%%%%%%%%%%%%%%%%%%%%%%%%%%%%%%%%%%%%%%%%%%%%%%%%%%%%%%%%%%%%%%%%%%%%%%%%%%%%%%%%%%%%%%%%%%%%%%%%%%%%

%%%%%%%%%%%%%%%%%%%%%%%%%%%%%%%%%%%%%%%%%%%%%%%%%%%%%%%%%%%%%%%%%%%%%%%%%%%%%%%%%%%%%%%%%%%%%%%%%%%%%%%%%%%%%%%%%%%%%%%%%%%%%%%%%%%%%%%%%%%%%%%%%%%%%%%%%%%%%%%%%%%%%%%%%%%%%%%%%%%%%%%%%%%%%%%%%%%%%%%%%%%%%%%%%%%%%%%%%%%%%%%%%%%%%%%%%%%%%%%%%%%%%%%%%%%%%%%%%%%%%%%%%%%%%%%%%%%%%%%%%%%%%%%%%%%%%%%%%%%%%%%%%%%%%%%%%%%%%%%%%%%%%%%%%%%%%%%%%%%%%%%%%%%%%%%%%%%%%%%%%%%%%%%%%%%%%%%%%%%%%%%%%%%%%%%%%%%%%%%%%%

\end{document}